\documentclass[journal=jpcbfk,manuscript=article,layout=twocolumn]{achemso}
\SectionNumbersOn

\usepackage{amsmath}

\newcommand*\etal{\emph{et\,al.}}
\newcommand*\kjmol{kJ\,mol$^{-1}$}
\newcommand{\B}{\begin{eqnarray}}
\newcommand{\E}{\end{eqnarray}}
\newcommand{\dd}{\, \mathrm{d}}

\author{Michal H. Kol\'a\v{r}}
\email{michal.kolar@uochb.cas.cz}
\affiliation[Forschungszentrum J\"{u}lich]{Institute for Advanced 
Simulations (IAS-5) and Institute of Neuroscience and Medicine (INM-9), 
Forschungszentrum J\"{u}lich, 52428 J\"{u}lich, Germany}
\affiliation[Institute of Organic Chemistry and Biochemistry]{Institute
of Organic Chemistry and Biochemistry
of the Czech Academy of Sciences, 16610 Praha, Czech Republic}
\altaffiliation{Current address: Department of Theoretical and
Computational Biophysics, Max Planck Institute for Biophysical
Chemistry, Am Fassberg 11, 37077 G\"ottingen, Germany}
\author{Tom\'a\v{s} Kuba\v{r}}
\email{tomas.kubar@kit.edu}
\affiliation[Karlsruhe Institute of Technology]{Institute of Physical 
Chemistry \& Center for Functional Nanostructures,
Karlsruhe Institute of Technology, 76131 Karlsruhe, Germany}

\title{Reaction Path Averaging: Characterizing Structural Response
of the DNA Double Helix to Electron Transfer}

\begin{document}

\begin{abstract}
A polarizable environment, prominently the solvent, responds to 
electronic changes in biomolecules rapidly.
The knowledge of conformational relaxation of the biomolecule itself,
however, may be scarce or missing.
In this work, we describe in detail the structural changes in DNA 
undergoing electron transfer between two adjacent nucleobases.
We employ an approach based on averaging of tens to hundreds of thousands
of non-equilibrium trajectories generated with molecular dynamics
simulation, and a reduction of dimensionality suitable for DNA.
We show that the conformational response of the DNA proceeds along
a single collective coordinate that represents the relative 
orientation of two consecutive base pairs, namely a combination of
helical parameters shift and tilt. The structure of DNA relaxes on 
time scales reaching nanoseconds, contributing marginally to the relaxation
of energies, which is dominated by the modes of motion of the aqueous
solvent. The concept of reaction path averaging (RPA), conveniently 
exploited in this context, makes it possible to filter out any 
undesirable noise from the non-equilibrium data, and is applicable
to any chemical process in general.
\end{abstract}



\section{Introduction}

Obtaining microscopic structural information about chemical reactions
is challenging due to their transient character. They are simply too
fast to be visualized conveniently. Also, due to stochastic motions at
a finite temperature, it might not be obvious which structural modes, or
collective coordinates are subject of major changes, and which remain
rather unaffected. The simplest chemical reaction is a transfer of a single
electron between two well-defined chemical species. Those can be separated
ions as in the initial Marcus' studies \cite{Marcus56}, or distinct 
chemical groups within a single (bio)macromolecule \cite{Marcus85}. 
Electronic changes in biomolecules have attracted particular attention
due to their role in metabolism as well as their connection to
urgent topics such as cancer or cell aging.
For instance, upon electromagnetic irradiation, the deoxyribonucleic acid
(DNA) undergoes processes that may eventually lead to changes in its 
chemical structure \cite{Lodish12}. These structural changes, albeit 
local, affect biochemical and mechanical properties of the DNA double
helix on a global scale \cite{Burrows98, Sinha02, Hada08}. This often has
profound consequences to its biological function, driving the cell to 
erroneous states \cite{DeBont04}.

The nucleobases are the most photosensitive parts of the DNA, 
therefore great efforts have been taken to describe their electronic
properties, regarding both excitation and 
ionization \cite{Crespo04, Pluharova15, Improta16}. An electron hole
created through oxidation of a nucleobase may travel along the double
helix to large distances of several dozens of
nanometers \cite{Hall96, Schuster04}.
Among the components of which DNA is built, purine nucleobases
possess the lowest ionization potentials, thus the transfer 
of an electron occurs primarily along guanines and 
adenines \cite{Giese01}. Interestingly, such electron-transfer
ability may allow to use DNA as a wire in nanotechnology 
applications, attempts of which have already 
appeared \cite{Guo16, Teschome16}. 

Apart from ionization, electromagnetic irradiation of DNA may also lead to
an excitation of nucleobases or to the formation of an excimer/exciplex
\cite{Middleton09}.
Then, the excitation energy may be transferred within the DNA molecule
\cite{Nordlund1993, Markovitsi2007, Vaya12}.
This process is most likely very complex as indicated by the growing
evidence that the excitation within DNA is delocalized over two
\cite{Crespo05} or more \cite{Kwok06, Buchvarov07, Huix15} nucleobases.

A change of the electronic structure of a biomolecule is accompanied
by a complex response of the aqueous environment. In return, such 
a modulated hydration dynamics affects the processes within 
the biomolecule, thus constituting an important functional 
aspect \cite{Fogarty13}. Water is perceived as a solvent 
possessing dynamics on a wide range of time scales. The fastest
one of a few tens to hundreds of femtoseconds is driven by 
rotational (librational) motions of the first solvation shell,
whereas slower motions at a picosecond scale involve concerted
angular jumps of hydrogen bonds between several water 
molecules \cite{Maroncelli88, Laage06, Rey15}.

Time-resolved Stokes shift (TRSS) and related experiments suggest
that the dynamics of water slows down in the vicinity of a hydrophilic
solute such as DNA \cite{Andreatta05, Berg08, Laage11}, presumably due
to a coupling with the solute degrees of freedom. In a classic setup
of TRSS experiments on DNA, a dielectric response to the de-excitation
of a fluorescent reporter bound to DNA is measured. Sen \etal ~provided
a robust interpretation\cite{Sen09} of their data covering the time 
scales ranging from 40 fs to 40 ns \cite{Andreatta05}, arguing that the 
largest part of the response is caused by water. On the other hand,
the ions present in solution and the DNA itself were identified to 
play minor roles. It seems, however, that these experimental efforts
were, on their own, unable to unravel any distinctive time scales 
and characteristic motions.

Notwithstanding, other studies provided at least partial information
of this kind. In a series of experiments with femtosecond resolution,
the nucleobase analogue 2-aminopurine was used as a chromophore
by Zewail \etal ~Two well-separated decay times of hydration were 
observed, 1.0 and 10--12 ps \cite{Pal03a}, and the solvation 
dynamics was slower in a drug$\cdots$DNA complex, with a decay
time of ca.\ 20~ps \cite{Pal03b}. Existence of two general kinds
of solvent molecules was inferred, and the slower was termed 
``dynamically ordered water''. In addition, an even faster 
component ($< 50$~fs) was concluded to dominate solvation
dynamics, in a combination of experimental and 
computational efforts over 20 years ago\cite{Jimenez94}.

Molecular dynamics (MD) simulations have become a valuable
and convenient complement to experimental investigation of
dynamics of such complex systems. Previous MD studies brought
insight into the energetics and structural relationships of 
the electron transfer within DNA \cite{Voityuk04, Kubar08}.
Proceeding further, a non-adiabatic
quantum-chemical/molecular-mechanical (QM/MM) scheme
for direct simulation of electron transfer in DNA and other
biomolecular complexes \cite{Kubar10, Kubar13} was constructed.

Concerning simulation studies related to TRSS experiments,
Pal \etal ~found that the DNA contributes little to the TRSS 
response, which is dominated by the solvent \cite{Pal06a, Pal06b}.
Sen \etal ~argued more specifically that ``All of the 
subpicosecond dynamics [leading to a simulated TRSS response] 
can be attributed to the water,'' and they did not find any 
important DNA contribution either \cite{Sen09}.
By contrast, another study identified the longer of 
the two time scales (20 ps) with movements of DNA 
clearly \cite{Furse08}. This discrepancy was discussed 
later\cite{Sen09}, and it was pointed out that the binding pose
of a low-molecular ligand complexed with DNA may affect 
the measured decays: unlike intercalating ones, minor-groove
binding molecules replace water molecules, which may affect 
the hydration pattern around the nucleobases modulating the 
TRSS response eventually. This makes the comparison of 
experiments performed on pristine
DNA \cite{Pal06a, Pal06b}, DNA$\cdots$intercalator
complex \cite{Sen09} and DNA$\cdots$minor-groove binder 
complex \cite{Furse08} somewhat complicated.

Another point worth mentioning is that the TRSS decays, which
are inherently non-equilibrium events, were modeled on the basis
of equilibrium simulations in most of these previous studies,
by way of a linear response approximation. The reason was
that ``although [the non-equilibrium simulation based quantity]
is more directly related to [TRSS] experiments, it requires 
an enormous amount of sampling to achieve good statistics
for quantitative analysis,'' as argued by Furse and 
Corcelli \cite{Furse08}.

To the best of our knowledge, no structural information about the
electron transfer in DNA is currently available. It does not seem to be
known if and how the double-stranded DNA deforms, and the structural
relaxation to the new equilibrium state has not been described yet.
Several proposals for the mechanism of electron transfer in DNA
involve special structural arrangements of
nucleobases \cite{ONeill04, Zhang15} but a solid proof is missing.
Therefore, the first aim of this work is to identify the modes
of motion of DNA that respond to the electron transfer,
and to estimate the time scales of such structural
relaxation. Since the changes of electronic structure are 
concentrated in nucleobases, emphasis is laid on the 
configuration of base pairs, and the structure of DNA is 
described with appropriate collective 
coordinates \cite{Dickerson89}.

\begin{figure}[tb]
\includegraphics{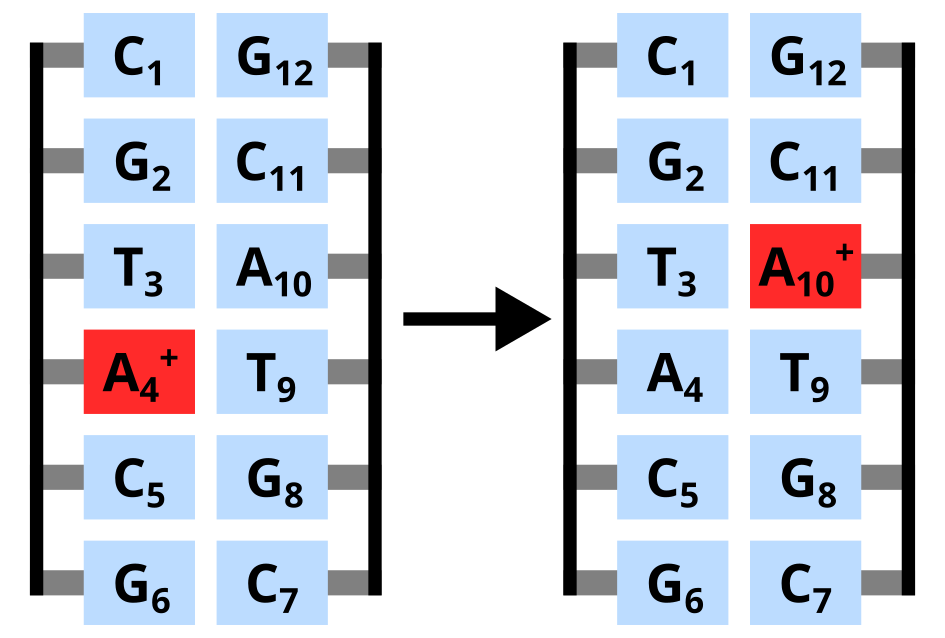}
\caption{Schematic representation of the electron transfer studied.
A radical cationic adenine is shown in red, whereas neutral ground state 
nucleobases are in blue. The equivalent process involving A$_{10}^+$ in the
initial state is also studied.}
\label{fig:reactions}
\end{figure}

This work deals with a small model molecule, a double-stranded
DNA hexanucleotide with a self-complementary sequence d(CGTACG)$_2$.
We investigate the transfer of an electron between
the two central adenines (Figure~\ref{fig:reactions});
note that the adenines in the ((T$_3$, A$_{10}$), (A$_4$, T$_9$))
base-pair step (here denoted as TA step) exhibit a large electronic
coupling,\cite{Kubar08} making the electron transfer efficient. 
An additional piece of motivation to use a palindromic DNA sequence
is that the forward and backward electron transfer reactions are actually
identical processes. This makes it possible to assess convergence
with a simple comparison of results for the forward and backward transfer.

It should be emphasized that the structural relaxations occur
on a rugged energy landscape in a multi-dimensional 
configuration space, which makes 
the visualization and the microscopic understanding of such 
processes difficult. Hence, we model the real non-equilibrium
processes by means of non-equilibrium MD simulations, similarly 
to the studies of vibrational relaxations in pure liquid
water\cite{Ingrosso09, Petersen13}, or of protein 
quakes\cite{Brinkmann16}. The initial and final
states may be connected via many reaction paths, and the high 
efficiency of today's computer hardware and software makes it 
possible to generate abundance of non-equilibrium trajectories
to sample the space of the reaction paths. This way, the
undesired noise is filtered out by averaging of the trajectories
in the space of suitable collective coordinates. The slow 
convergence of results that are to be averaged is accepted as 
an unfortunate fact, which is under full control, however. 
The technique is referred to as Reaction Path Averaging (RPA)
throughout the paper, and its description is another aim of 
this work.

The paper is organized in two major parts dealing with the 
equilibrium and non-equilibrium simulations. Each of the parts
starts with a description of the methodology used, and 
follows with the outcome of the respective analyses. 
In the final section, we conclude with a discussion of the
results in the context of the previously reported work.


\section{Equilibrium Ensembles}
\label{sec:equil}

\subsection{Methods of Equilibrium Simulations}

The DNA oligonucleotide was described with the molecular mechanics (MM)
force field Amber parm99bsc0 \cite{Cornell95, Perez07}. 
The MM models of adenine in the radical cationic state was created by
adjusting the Amber partial atomic charges of the ground state,
to represent the electrostatic potentials obtained at the HF/6-31G* level
via restricted electrostatic potential fitting (RESP) \cite{Bayly93}
performed with the Antechamber module of AmberTools \cite{Amber14}.
Such an application of RESP is compatible with the Amber
force field. Quantum chemical calculations were performed with
Turbomole 6.5 \cite{Turbomole1,Turbomole2}.
The difference between the adjusted cationic and ground state charges
is sketched in Figure~S1. The electronic change involves almost all of the
adenine atoms.

Bond, angle, torsional and Lennard-Jones (LJ) parameters of the charged
state were considered equal to their neutral ground state values.
The use of ground-state LJ parameters is justified by the small deviations
of Bader's atomic volumes \cite{Bader87, Tang09} calculated
at the HF/TZVP level.
The root-mean-square deviations (RMSD) of atomic volumes from those
in the ground state amount to 13~\%.
The largest deviation was found for one of the amino hydrogens (33~\%).
Such changes of atomic volumes would translate into changes of the LJ radii
of at most 10~\%, corresponding to ca.\ 0.2~\AA{};
this is the maximal possible magnitude of differences of LJ radii
that was neglected.

Regarding the bond, angle and torsional parameters,
several reasons led us to keep them unchanged upon the electronic change.
First, the energy minimized geometries of the neutral ground and the
radical cationic states are almost identical at the B3LYP/6-31G* level;
the RMSD of heavy-atom coordinates is as low as 0.03~\AA{}.
Second, it is unclear if and how the quantum effects on the distribution
of bond lengths should be modeled in classical simulations.
A rather typical treatment of covalent bonds in classical MD simulations,
adopted also in the current work is to constrain their lengths
to the respective force-field equilibrium values \cite{Hess08},
even though this seems to be less usual in the field of DNA simulation.
Further, any structural analyses are performed in terms of collective 
coordinates, where nucleobases are represented as rigid bodies;
the internal degrees of freedom are filtered out in any case.
Finally, the aim is to describe mostly the interaction of a nucleobase
with its environment, for which the intra-nucleobase degrees of freedom
play little role.

The double-stranded DNA hexanucleotide d(CGTACG)$_2$ was immersed
in a periodic box filled with ca.\ 3000 SPC/E water 
molecules \cite{Berendsen87}, and the system was neutralized by addition
of ten sodium ions \cite{Joung08}, or nine whenever a radical cationic 
adenine was involved. The conformational space of the oligonucleotide 
with the radical cationic A$_4^+$ or A$_{10}^+$,
and the neutral ground state (GS) adenines was sampled by means
of ten independent equilibrium molecular dynamics (MD) simulations
of at least 510~ns each, for every of the different DNA species.
We note that the structure of an ionized DNA species 
may not be able to reach thermodynamic
equilibrium due to the short lifetime of the cationic state.
Still, the assumption that the cationic state lives long enough to 
make structural equilibration possible is a necessary one,
and present in all of the cited previous simulation work.
All of the simulations were performed with
Gromacs 4.6 \cite{Pronk13, Pall14}.
The entire simulation setup is further detailed in the Supporting
Information (SI).

In summary, the structural ensembles were generated with MD simulations
in which bond lengths were constrained to their equilibrium values,
while all of the remaining degrees of freedom including angles and
dihedral angles were free to vary.
For the purpose of any following structural analysis,
rigid-body alignment of nucleobases  was performed.
Thus, intra-molecular degrees of freedom were integrated out effectively
and were not analyzed.
DNA structures were described in terms of the helical parameters
within the standard reference frame as defined by Tsukuba conventions \cite{Olson01},
which were calculated with the 3DNA program suite \cite{Lu08} interfaced
to Gromacs by means of the do\_x3dna utility \cite{Kumar15}.
The numpy \cite{VanDerWalt11} and matplotlib \cite{Hunter07} libraries
were utilized in most of the python analyses.

\subsection{Equilibrium Structure}

The relatively high rigidity of nucleobases allows the atomistic
representation of the DNA structure to be reduced into a few relevant
coordinates \cite{Dickerson89}, here referred to as the helical 
parameters. They describe the relative orientation either of the
nucleobases within a base pair, or of two consecutive base pairs,
i.e.\ the base-pair step (see Figure~S2).

We make use of the concept of helical parameters to describe
the structure of the central TA step, which is involved in the 
electron transfer reaction investigated.
Regarding the equilibrium states, mean values of the
helical parameters over the equilibrium ensembles were
calculated to infer the structural change of the central 
TA step associated with the electron transfer reaction.
Note that mostly the mean values are
discussed, and the variance and statistical uncertainty
expressed in terms of the respective standard deviations 
and standard errors of the mean are presented in the SI. 
Apart from the helical parameters, the dihedral angles in the
sugar-phosphate backbone and the sugar puckering
were analyzed as well.

The equilibrium structure of the central TA step with both adenines
uncharged (in the neutral ground state)
is in between the A- and B-DNA conformations;
all of the helical parameters are summarized in Tables~S2 and S3.
In comparison with the sequence-averaged data from high-resolution crystal 
structures \cite{Olson01}, the values of roll and twist
are closer to A-like values, whereas the other helical 
parameters resemble a B-like conformation more closely (Table~S3).
While it is a known deficiency of current DNA force fields that they 
underestimate twist, the equilibrium values of twist lie
within the error bars of both A-DNA and B-DNA here.
The distributions of helical parameters are close to normal, see
Figures~S3--S4, with an exception of the slide, which hints at a bimodal
distribution. We note that no particular DNA conformation 
is \emph{a priori} assumed in any of the following discussions,
and the conformational changes are rather described by the actual
values of helical parameters.

We also analyzed the structure of the sugar moieties of the four
nucleotides composing the TA base-pair step,
see Tables~\ref{tab:b1b2}, \ref{tab:pucker}, S5 and S6.
The canonical BI substate is dominant, with a propensity
of 85~\% and 96~\% for the adenosines and the thymidines, respectively,
and a mean lifetime of ca.\ 400~ps.
The mean lifetime of the BII substate is ca.\ 70~ps and ca.\ 20~ps
for the adenosines and the thymidines, respectively.
Note, however, that the BI/BII equilibrium remains an open issue in general: 
``A reliable experimental estimate of the BII population 
remains so far a major challenge\ldots{} Additional work 
is required to clarify the issue.''\cite{Drsata12}. 
The sugar puckering is rather diverse, with the canonical 
C2'-endo substate populated only to 33~\% and 17~\% in each
of the two adenosines and each of the two thymidines, respectively.
All of the sugar conformations are relatively short-lived,
with mean lifetimes below 4~ps.
A positive charge located on one of the adenines stabilizes the BI substate
of its own backbone, whereas it destabilizes BI of the opposite strand;
the propensity of BI changes from 85~\% in the ground state
to 99~\% and 45~\%, respectively (Tables~\ref{tab:b1b2} and S5). 
Likewise, the mean lifetime of BI of the charged adenosine increases to 3.5~ns,
and that of the opposite adenine decreases to 165~ps (Table~S5).

\begin{table*}
\begin{tabular}{lrrrr}
\hline
    &  T$_3$  &  A$_{10}$  &  A$_4$   &  T$_9$ \\
\hline
GS         &  96 &  86 &  85 &  96 \\
A$_4^+$    &  32 &  44 &  99 &  96 \\
A$_{10}^+$ &  96 &  99 &  47 &  31 \\
\hline
\caption{Populations (in \%) of the BI substate of the four
nucleotides in the TA base-pair step. BI is considered
whenever the difference of backbone dihedral angles
$\varepsilon - \zeta < 0$. The remainder to 100 \%
pertains to the BII substate.}
\label{tab:b1b2}
\end{tabular}
\end{table*}

\begin{table*}
\begin{tabular}{lrrrr}
\hline
     &     T$_3$ & A$_{10}$ &  A$_4$ &   T$_9$ \\
\hline
   C1'-exo &    36 &    31 &    32 &    35 \\
  C2'-endo &    17 &    34 &    33 &    17 \\
   C3'-exo &     1 &    10 &    10 &     1 \\
   C4'-exo &    12 &     6 &     6 &    13 \\
  O4'-endo &    32 &    17 &    17 &    33 \\
\hline
\caption{Populations (in \%) of various conformations 
of the deoxyribose ring (i.e., sugar pucker) of the 
nucleotides in the TA base-pair step in the neutral ground electronic state.
These data for DNA with A$_4^+$ or A$_{10}^+$ are presented in 
Table~S7.}
\label{tab:pucker}
\end{tabular}
\end{table*}

\begin{figure}[tb]
\includegraphics{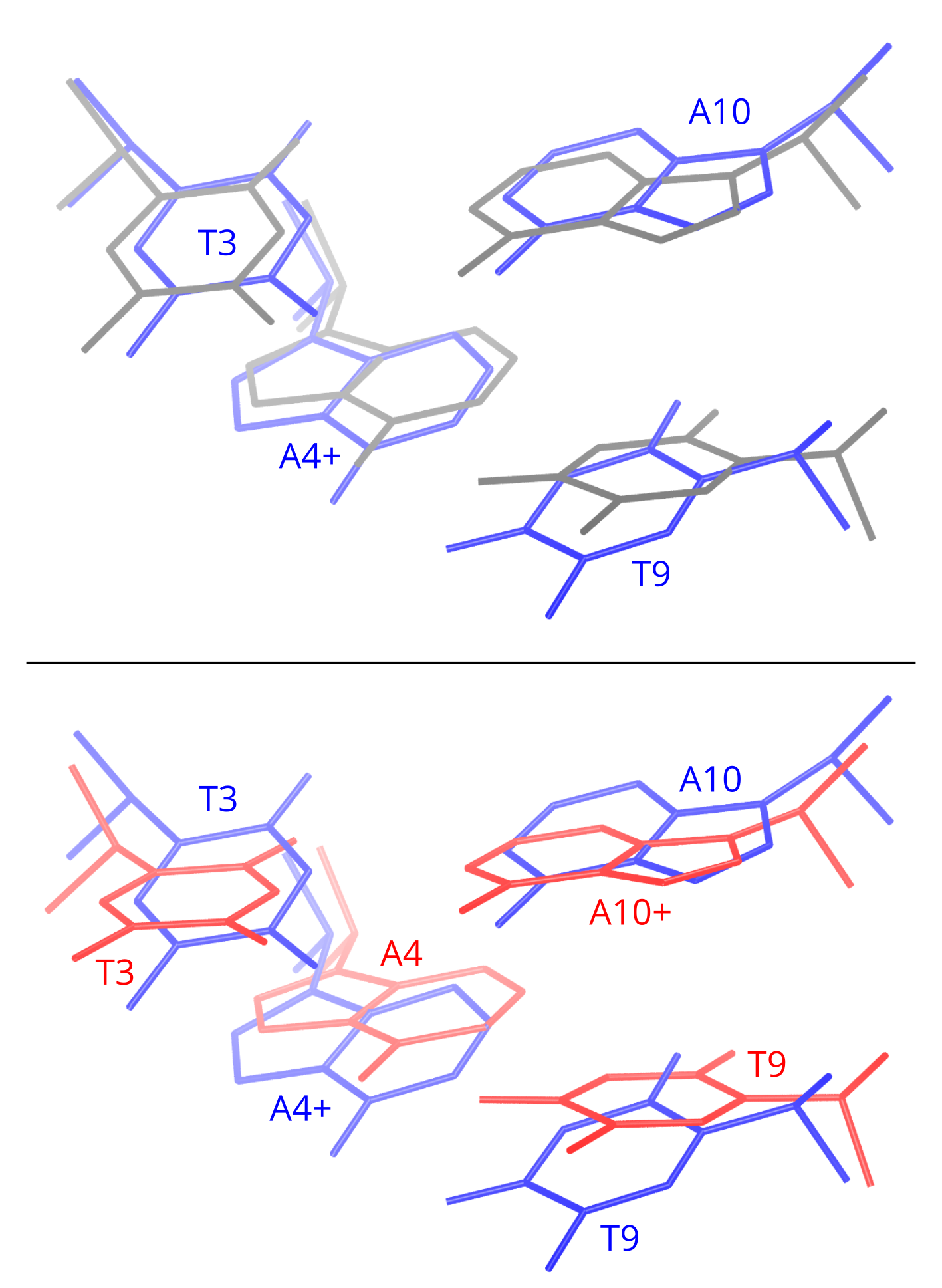}
\caption{The average structures of the TA step with both adenines
in the neutral ground state (gray), radical cationic A$_4^+$ (blue),
and radical cationic A$_{10}^+$ (red). Hydrogen atoms are omitted 
for clarity.}
\label{fig:structures}
\end{figure}

The structural changes upon electron transfer are rather minor
at the level of individual base pairs  (Table~S2). 
Merely the changes of propeller and opening are
non-negligible, amounting to 7.5$^{\circ}$ and 5.5$^{\circ}$,
respectively, and reflecting well the sequence symmetry.
For the global DNA conformation, however,
the inter-nucleobase orientation is less important
than the relative orientation of consecutive base pairs. 
Thus, we turn our attention to the changes of base-pair 
step parameters of the central TA step instead,
and these are summarized in Table~\ref{tab:stepparams}.
%
The electron transfer is accompanied by a huge change of
3.4~\AA{} in shift, which stands for a relative movement
of two neighboring base pairs in the plane perpendicular to 
the helical axis in the direction from/to major/minor groove
(Figure~\ref{fig:helparams}). This conformational change
is seen clearly in the DNA structural models in Figure~\ref{fig:structures}
constructed from the mean values of the helical parameters (Tables~S2 
and S3). The differences of slide and rise vanish.
Among the angular parameters, the change in tilt
is the most pronounced and amounts to 5.3$^{\circ}$.
The changes of roll and twist are close to zero. Hence, regarding
the relaxations, we focus our attention to the changes of shift 
and tilt. The good convergence of the equilibrium ensembles 
is reflected by symmetric values of the relevant parameters 
as shown in Table~\ref{tab:stepparams}. The standard errors 
of the mean are low, too (Tables~S2--S4).

\begin{figure}[tb]
\includegraphics{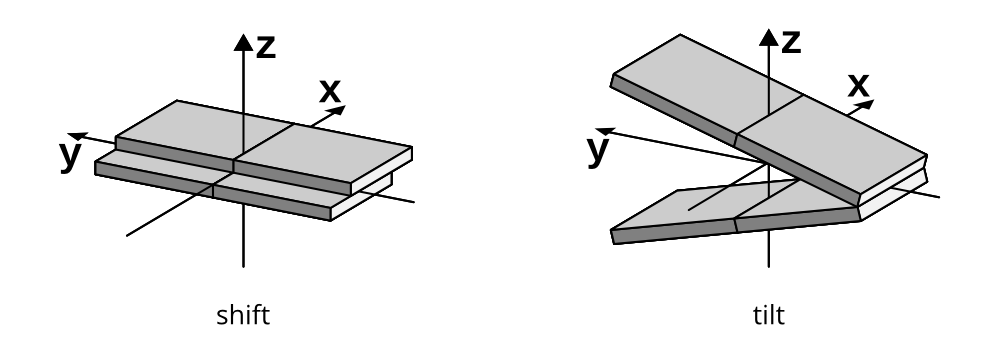}
\caption{The DNA base-pair step parameters shift and tilt. 
Each box represents one nucleobase.}
\label{fig:helparams}
\end{figure}

\begin{table*}
\begin{tabular}{lrcrcrc}
\hline
 & $\mathrm{mean\,}\pm\mathrm{\,std}$ & $\mathrm{sem} \times 10^3$ \\
\hline
$\Delta$shift &
$ 3.40\pm 1.25$ & 0.89 \\
$\Delta$slide &
$ 0.01\pm 0.73$ & 0.52 \\
$\Delta$rise  &
$ 0.01\pm 0.42$ & 0.30 \\
$\Delta$tilt  &
$ 5.32\pm 7.16$ & 5.10 \\
$\Delta$roll  &
$-0.11\pm 9.74$ & 6.94 \\
$\Delta$twist &
$ 0.46\pm 8.34$ & 5.93 \\
\hline
\end{tabular}
\caption{
The overall change in helical parameters of the central TA 
base-pair step
during the reaction $\mathrm{A}_4^+\rightarrow\mathrm{A}_{10}^+$
as represented by differences of mean values 
(\emph{mean}) calculated from the equilibrium ensembles 
equivalent to at least 5.1~$\mu$s. The variance and statistical
uncertainty are quantified by means of the standard deviation
(\emph{std}) and the standard error of the mean (\emph{sem}),
respectively. Shift, slide and rise in \AA; tilt, roll and 
twist in degrees. Note that the reaction $\mathrm{A}_{10}^+\rightarrow
\mathrm{A}_{4}^+$ yields values that are the negative of those
in the table due to 
symmetry. Equivalent data for the base-pair parameters are in
Table~S8. For the absolute values of the base-pair 
step parameters obtained for equilibrium
structural ensembles, refer to Tables~S3 and S4.}
\label{tab:stepparams}
\end{table*}

Also interesting is to compare the above observations with
the structural dependence of the electronic coupling between nucleobases,
which is one of the determinants of electron transfer.
Voityuk \etal{}\cite{VoityukPCCP01} reported a strong dependence
of electronic coupling on rise and slide, but almost none on shift and tilt,
for the ((TA), (AT)) base-pair step involving uncharged nucleobases only.
This is difficult to correlate with our findings above,
thus no connection is apparent between the structural relaxation of DNA
upon electron transfer and structural dependence of the electronic coupling.

\subsection{Equilibrium Energetics}

Motivated by the relation to TRSS experiments, the previous 
simulation studies concentrated mostly on the energy relations
in the DNA molecule, between DNA and the solvent, and their 
response to the de-excitation of a component of the 
system (reviewed by Furse and Corcelli).\cite{Furse10} The
equilibrium trajectories generated here may be analyzed to 
estimate the magnitude of energy that has to relax after an
electron transfer event. This is performed by evaluating the
MM component of the diabatic energy gap (DEG),
\begin{align}
\mathrm{DEG}(t=0) & =  \left < E(2) \right > _1 -
\left < E(1) \right > _1 = \nonumber \\
                  & = \left < E(2) - E(1) \right > _1,
\label{eq:degini}
\end{align}
for the initial state (time $t=0$) of a process 1$\rightarrow$2,
where $E(x)$ is the total potential energy evaluated with the 
Hamiltonian of state $x$, and $\left<\,\right>_y$ represents
the averaging of energy over an equilibrium ensemble generated 
with the Hamiltonian of state $y$. DEG for the final state of 
that process follows in analogy,
\begin{align}
\mathrm{DEG}(t=\infty) & =  \left < E(2) \right > _2 - 
\left < E(1) \right > _2 \nonumber \\
                       & = \left < E(2) - E(1) \right > _2.
\label{eq:degfin}
\end{align}

Generally, DEG is a measure of the relative magnitude of
intermolecular interactions exercised by two different electron
distributions with the remainder of the same molecular structure.
In the case of electron transfer with vanishing reaction energy, 
the DEG would correspond to the outer-sphere reorganization 
energy $\lambda$ in the context of Marcus' theory of electron
transfer \cite{Marcus85}. Note that the reorganization energy
for the process $\mathrm{A}_4^+\rightarrow\mathrm{A}_{10}^+$,
for instance, is the difference of energy evaluated with 
the Hamiltonian of A$_4^+$ for the ensemble of structures
of A$_{10}^+$ and that of A$_4^+$,
\begin{equation}
\lambda = \left < E(\mathrm{A}_4^+) \right > _{\mathrm{A}_{10}^+}
- \left < E(\mathrm{A}_4^+) \right >_{\mathrm{A}_4^+}.
\end{equation}

The values of DEG obtained for the initial and final states
of the considered reactions are shown in Table~\ref{tab:deg}.
DEG for the reactant and product states of the electron 
transfer processes amounts to 177~\kjmol{} (with the 
appropriate sign), and this value is in agreement with our
previously reported reorganization energies obtained with 
a similar methodology for similar DNA species \cite{Kubar09}.
Note that if the purpose of these calculations was to estimate 
reorganization energies, the values should be scaled down by
a factor of ca.\ 1.4 to correct for the neglect of electronic
polarization in classical MD simulations \cite{Tipmanee10}.
A result that is more relevant for the current work, though,
is the change of DEG when passing from the initial to the 
final state of each respective reaction considered.
In the electron transfers, DEG passes from 177~\kjmol{} to
--177~\kjmol{}, i.e.\ it decreases by 354~\kjmol{}.

In the following, DEG as a function of time is used
to probe the response of interactions in the DNA-solvent
system during the relaxation simulations.

\begin{table}
\begin{tabular}{lcc}
\hline
~& $\mathrm{mean} \pm \mathrm{std}$ & sem \\
\hline
DEG(0) & $176.4\pm540.2$ & 1.48 \\ 
DEG($\infty$) & $-177.9\pm539.6$ & 1.45 \\
\hline
\caption{Diabatic energy gap for the initial state, DEG(0),
and for the final state, DEG($\infty$), of the electron
transfer $\mathrm{A}_4^+ \rightarrow \mathrm{A}_{10}^+$ obtained from the
equilibrium ensembles. Mean values (\emph{mean}), standard 
deviations (\emph{std}), and standard errors of the mean
(\emph{sem}) are provided.
Note that the reaction $\mathrm{A}_{10}^+\rightarrow
\mathrm{A}_4^+$ yields identical values due to 
symmetry.}
\label{tab:deg}
\end{tabular}
\end{table}

\section{Relaxations}
\label{sec:relax}

\subsection{Methods of Relaxations}
\label{sec:relax-methods}

\begin{figure}[tb]
\includegraphics{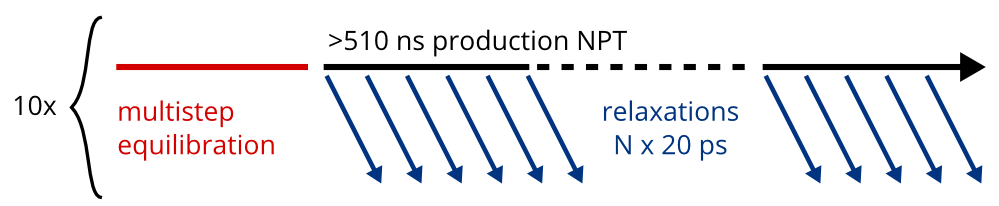}
\caption{Non-equilibrium MD simulation scheme.}
\label{fig:simulations}
\end{figure}

To investigate the time evolution
of the structural and energetic changes, relaxation simulations
of 20~ps each were started from the appropriate equilibrium 
ensembles, which had been created by taking snapshots from 
the multi-100ns equilibrium MD trajectories of the end states
in regular time intervals, as depicted in 
Figure~\ref{fig:simulations}. The changes of electronic 
structures were modeled by means of instantaneous changes
of the MM point charges of the adenines at time $t=0$. 
50,000 relaxations were simulated for each of the electron 
transfers $\mathrm{A}_4^+\rightarrow\mathrm{A}_{10}^+$
and $\mathrm{A}_{10}^+\rightarrow\mathrm{A}_4^+$.
To assess any slower response, additional relaxation simulations 
were performed: 500 simulations of 1 ns each, and 100 simulations
of 10 ns each. The resulting RPA time series of energies and of
helical parameters were obtained by averaging over the 100,000
shorter relaxations, as well as 1000 or 200 longer simulations,
respectively.

Each of the averaged data series was fitted with
a multi-exponential function in order to estimate the 
relevant decay times. To this end, denser data from 
20-ps-long simulations (with a time resolution of 2~fs)
were combined with sparser data from 1-ns-long simulations 
(after block averaging over 50 data points, leading to 
the resolution of 100~fs). First, the longest relaxation 
time $\tau$ was determined with an exponential fit, 
$f(t) = w \exp [-t/\tau]$ to the 250--1000~ps region of 
the data series. The relaxation time $\tau$ was kept fixed
then. In the second step, a more complex function composed
of several, optionally stretched exponentials was fitted 
to the entire data series (0--1000~ps). The following 
function was considered,
\begin{align}
f(t) = & a_1 \exp \left[ -\frac{t}{t_1} \right]
         + a_2 \exp \left[ -\left( \frac{t}{t_2} \right)^{\!\! c_2} \right] + \nonumber \\
       & + a_3 \exp \left[ -\left( \frac{t}{t_3} \right)^{\!\! c_3} \right]
         + a_4 \exp \left[ -\frac{t}{\tau} \right] ,
\label{eq:fit-fn}
\end{align}
with the second and third exponential optionally stretched.
The relaxations of the normalized energies (see below)
employed a boundary condition of $a_1 + a_2 + a_3 + a_4 = 1$.
The relaxation of DNA helical parameters was fitted with 
a triple exponential (and $a_3 = 0$);
the coefficients $a_i$ have the dimension of the helical
parameter being fitted (\aa{}ngstr\"{o}m for shift, degree 
for tilt), and their sum $a_1 + a_2 + a_4$ shall correspond
to the total magnitude of the relaxation shown in 
Table~\ref{tab:stepparams}. A comparison of the multiple
tested choices of optionally stretched exponentials can be 
found in the SI.

Prior to fitting, the baselines obtained from the equilibrium
simulations were used to shift all of the relaxation data 
series to zero at infinite time. All of the fitting was 
performed with the Levenberg--Marquardt non-linear 
least-squares algorithm \cite{Levenberg, Marquardt} as 
implemented in the R package.\cite{R,R-minpack.lm}
The initial guess of the parameters for fitting, and
the fitted values for the ten individual sub-ensembles 
for each relaxation are presented in the SI. For every 
fitted exponential component, a decay time $T_i$ was obtained.
For simple exponentials, $T_i$ equals directly to the parameter
$t_i$; for stretched exponentials, this is calculated as
$T_i = t_i / c_i \cdot \Gamma(1 / c_i)$ with the gamma function
$\Gamma(x) = \int_0^\infty y^{x-1} \exp [-y] \dd y$.
The accuracy of the obtained fits was judged by the ratio
of the residual sum of squares and the total sum of squares,
RSS/TSS. Its calculation is described in the SI, as is the
method to estimate the statistical uncertainty of
determination of the fitting parameters.

\subsection{Energy relaxation}
\label{sec:relax-degn}

We monitored the time course of energy relaxation with the 
time-dependent diabatic energy gap, which was obtained along 
the trajectory from every relaxation simulation 
1$\rightarrow$2 as $\mathrm{DEG}(t) = E(2,t) - E(1,t)$, 
where $E(x,t)$ is the potential energy obtained with the
Hamiltonian of state $x$ at time $t$ in the simulation. 
Then, these time series were averaged over the respective 
sets of relaxation simulations. Finally, DEG($t$) was 
normalized to the interval $(0,1)$ yielding
the normalized diabatic energy gap (DEGN):
\begin{equation}
\mathrm{DEGN}(t) = \frac{ \mathrm{DEG}(t) - \mathrm{DEG}(\infty)}
{ \mathrm{DEG}(0) - \mathrm{DEG}(\infty)}.
\label{eq:degtime}
\end{equation}
DEGN is very similar to the quantities employed previously
for electric potentials by Maroncelli and 
Fleming \cite{Maroncelli88} as well as for differential 
interaction energies by Sen \etal ~\cite{Sen09} and Furse
and Corcelli\cite{Furse08, Furse10}. An advantage of our
approach involving the non-equilibrium simulations is
the fact that the needed baselines DEG(0) and 
DEG($\infty$) were determined accurately in the preceding
equilibrium simulations, so that they need not be 
estimated with any, possibly difficult fitting. Note that 
the difficulties of determining the baselines with 
a sufficient accuracy constitute two of the ``three factors
that complicate the comparison of the experimental TRSS to
the simulated [quantities]'' within the linear response
approximation \cite{Sen09}.

\begin{figure}[tb]
\includegraphics{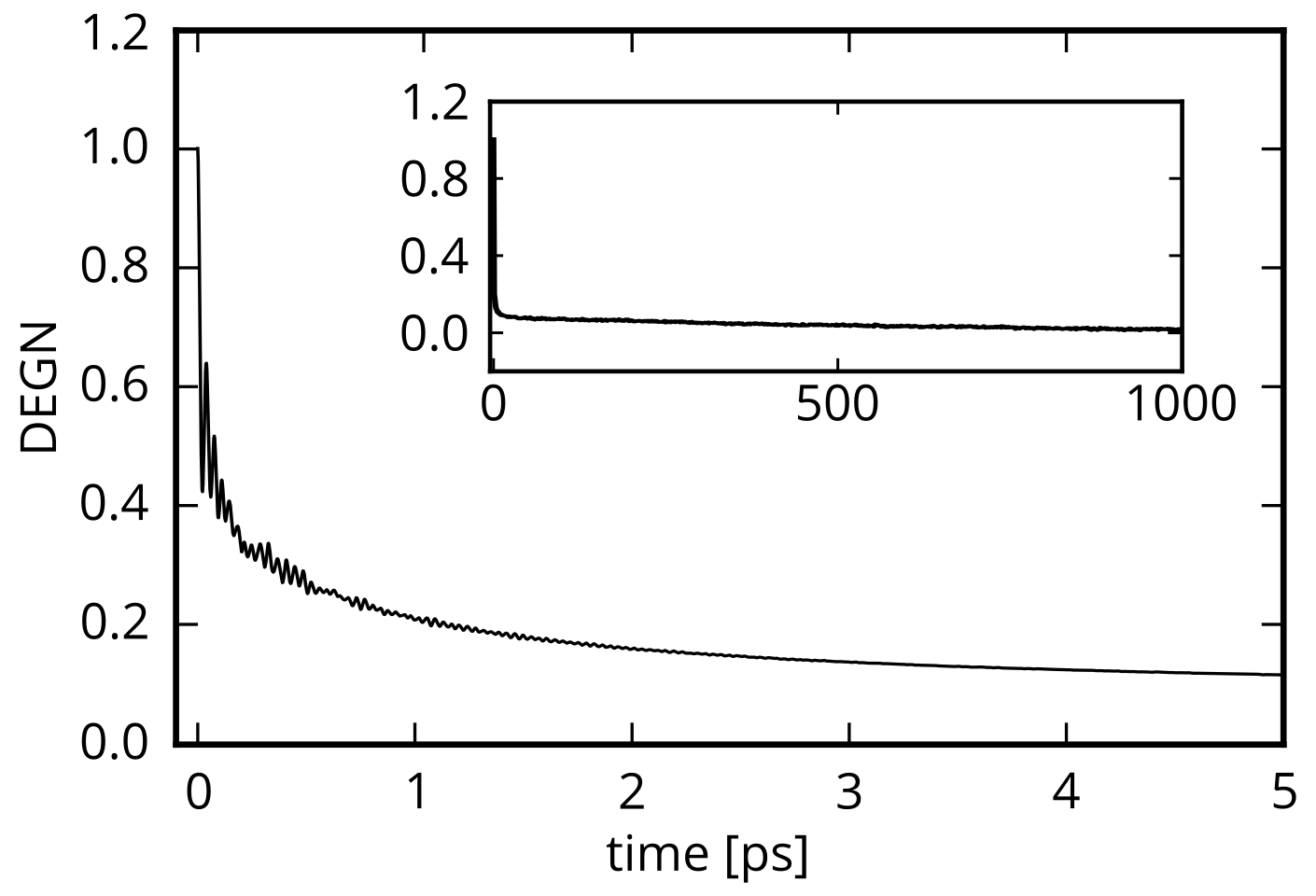}
\caption{Energy decay as captured by the normalized diabatic 
energy gap (DEGN) following an instantaneous electron
transfer. Data were obtained by averaging over the respective
numbers of non-equilibrium MD simulations, see the main text.
The inset shows the relaxation on a longer time scale.}
\label{fig:degn}
\end{figure}

The resulting averaged relaxation of DEGN is presented 
in Figure \ref{fig:degn}; note that every data point in this 
plot was obtained by averaging of 2$\times$50,000 values.
The initial phase of the decay is 
very steep, and about 80\% of the relaxation occurs within 
the first picosecond. The relaxation up to ca.\ 0.5~ps is 
apparently modulated by a periodic function with a small
amplitude and a period of about 45 fs, which corresponds to the 
wavenumber of 740 cm$^{-1}$. This frequency is in the range of
the librational motion of bulk water molecules \cite{Laage11},
Previously, such fluctuations were 
shown to induce fluctuations of instantaneous ionization
energies of DNA nucleobases \cite{Kubar08}.

No consensus seems to exist about what kind of analytic
function represents the response in the best way. For instance,
exponential, multi-exponential, stretched exponential or 
power-law functions have been proposed in the context of 
relaxations upon electronic changes in 
biomolecules \cite{Brauns02, Andreatta05, Pal06a, Conti10}. In this
work, a sum of two exponentials and two stretched exponentials
in Eq.~\ref{eq:fit-fn} provided an accurate fit of the 
averaged DEGN decay over the interval of 1~ns. It turned 
out that the periodic oscillations in the DEGN data
are too weak to be fitted, therefore no periodic component
was considered.

\begin{table*}
\footnotesize
\begin{tabular}{lcccccccccccccc}
\hline
          & $a_1$ & $t_1$ & $a_2$ & $t_2$ & $c_2$ 
          & $a_3$ & $t_3$ & $c_3$ & $a_4$ & $\tau$ & RSS/TSS \\
\hline
DEGN      & 0.34 & 0.0080 & 0.37 & 0.198 & 0.516 &
          0.21 & 1.43 & 0.488 & 0.08 & 654       & 0.0076 \\
uncert.   & 0.03 & 0.0004 & 0.11 & 0.052 & 0.090 &
          0.08 & 1.62 & 0.104 & 0.01 & 95        \\
\hline
\end{tabular}
\caption{Values of the parameters from the fitting of 
a multi-exponential function to the data series of DEGN
from the simulations of electron transfer;
$t_i$ in ps, dimensionless numbers otherwise. The accuracy of
the fit (RSS/TSS) and the estimates of uncertainty of the 
individual parameters were obtained as described in SI.}
\label{tab:degnrelax}
\end{table*}

\begin{table}
\begin{tabular}{lcc}
\hline
   & Magnitude & Decay Time  \\
   &  [\%]     &  [ps]       \\
\hline
1  &     34    &  0.0080     \\
2  &     37    & 0.374$^*$   \\
3  &     21    &  2.99$^*$   \\
4  &      8    &    654      \\
\hline
\multicolumn{3}{l}{$^*$ stretched exponential} \\
\end{tabular}
\caption{Relaxation times of DEGN and their relative 
magnitudes.}
\label{tab:degnrelaxtime}
\end{table}

The resulting parameters of the fitting function as well as
the summary of decay times and the corresponding decay 
magnitudes are presented in Tables~\ref{tab:degnrelax} 
and \ref{tab:degnrelaxtime}. First, there is an ultra-fast
component with a large magnitude and a decay time around 10~fs.
This agrees with the results from previous non-equilibrium
simulations by Furse and Corcelli,\cite{Furse08} who analyzed
a smaller number of more coarsely sampled trajectories,
while ``any sub-100-fs components in these dynamics [were]
unresolved''\cite{Pal03b} in typical TRSS experiments.
Then, there are two components with sizable magnitudes and
decay times of 0.4 and 3~ps. Lastly, there is a relatively 
weak component with a sub-nanosecond decay.

The experiments by Zewail \etal{}\ reported decay times of a nucleobase analogue
de-excitation of 1.4/1.0 and 19/10--12~ps (Ref.~\citenum{Pal03a}/Ref.~\citenum{Pal03b}).
The relaxations upon electron transfer are somewhat faster,
by factor of about three.
We note that such a comparison is not quite straightforward:
The decay in our simulations is a response to an electron transfer reaction. 
On the other hand, the experimental TRSS decays represent a response
to a de-excitation of a DNA component or of another molecule bound to DNA.
Importantly, recent reports suggested that the excitation/de-excitation of adenine 
is accompanied by a charge transfer from/to neighboring 
nucleobase(s) \cite{Middleton09,DoorleyJPCL13}, and that the electronic
re-arrangement upon (measured) de-excitation is delocalized just
like it is upon (simulated) electron transfer.
Therefore, just like the simulations in the current work, the TRSS experiments
involved a response to a charge transfer process,
albeit quantitatively different (i.e., with different quantitative characteristics
like spatial extent, driving force or reorganization energy).
The comparison of the current results to those from the TRSS experiments referred to above
is justified from this point of view.
In addition, to the best of our knowledge, no more closely related experimental data
are available.

The discussion of DEGN will be resumed in 
Section~\ref{sec:relax-helical} as soon as the relaxation of 
structural coordinates is presented.

\subsection{Structural Relaxation}
\label{sec:relax-helical}

To characterize the relaxation of DNA structure upon
electron transfer, the base-pair 
and base-pair step parameters were monitored along each of 
the relaxation trajectories, and these time series were 
averaged over both sets of trajectories. Owing 
to the symmetry of the self-complementary DNA sequence,
the two reactions investigated are equivalent,
$\mathrm{A}_4^+ \to \mathrm{A}_{10}^+$
and $\mathrm{A}_{10}^+ \to \mathrm{A}_4^+$. The corresponding
relaxation simulations describe the same physical phenomenon,
and must lead to the same results. Therefore, the time series
obtained from the respective pairs of relaxation simulations
were combined prior to analysis in order to improve the 
signal-to-noise ratio.

\begin{figure*}[tb]
\includegraphics{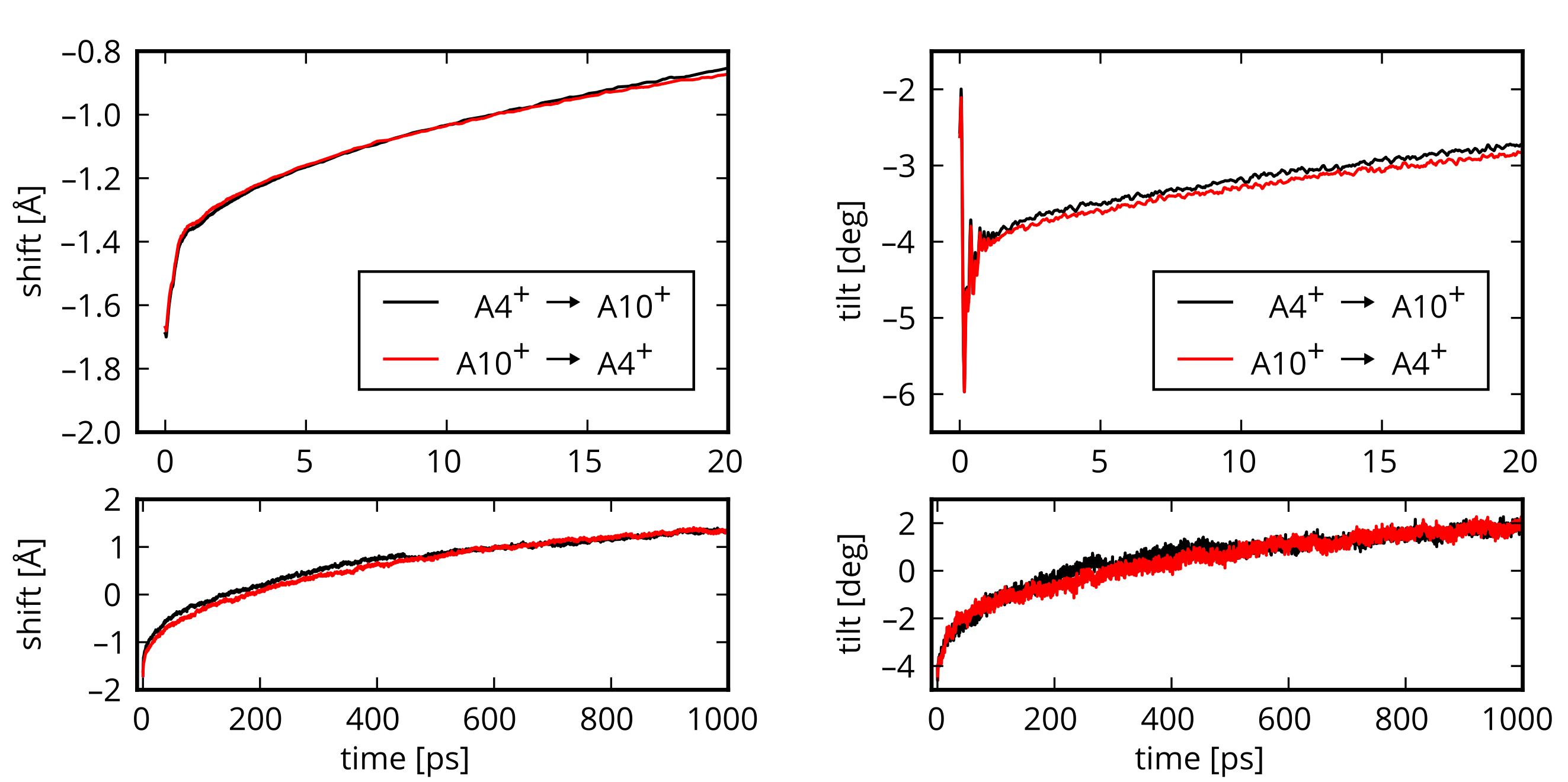}
\caption{Time evolution of the base-pair step parameters shift
and tilt. Note that the actual values for reactions 
$\mathrm{A}_4^+\rightarrow\mathrm{A}_{10}^+$
and $\mathrm{A}_{10}^+\rightarrow\mathrm{A}_4^+$ have opposite 
sign owing to the symmetry of the DNA sequence and the 
anti-symmetry of shift and tilt. For clarity, the values of 
parameters for $\mathrm{A}_{10}^+\rightarrow\mathrm{A}_4^+$
were inverted before plotting.}
\label{fig:structrelax}
\end{figure*}

Focusing on the most dynamical structural parameters as 
identified by the equilibrium ensembles, the relaxations of
shift and tilt are shown in Figure~\ref{fig:structrelax};
refer to Figure~S6 for the full set of relaxation profiles.
Immediately after the instantaneous electron transfer 
at $t=0$, within the first picosecond,
both shift and tilt as well as all of the other parameters
undergo strong fluctuations due to the sudden increase of 
energy in the system and its consequent redistribution.
This immediate response of the DNA is followed by a slower
relaxation, directed towards the final parameter values 
predicted from the equilibrium simulations.

Because of the symmetry of the considered palindromic
DNA sequence, the black and the red curves in 
Figure~\ref{fig:structrelax} should be identical. This is not
entirely the case; the differences of shift of up to
0.02~\AA{} between the two instances are observed.
Such a difference illustrates the extremely slow convergence
of the non-equilibrium data with respect to the number of 
simulations performed. Still, the shape of the pairs of curves
is remarkably similar, especially on the short time scale. 

The decays of shift and tilt were subject of a fitting 
procedure as detailed above. A sum of two exponentials and
one stretched exponential provided the best fit.
The initial phase of the relaxation was particularly difficult
to fit, and there were especially large oscillations in the case
of tilt at the start of the electron transfer relaxation 
simulations (see Figure~\ref{fig:structrelax}). These 
oscillations brought the averaged tilt from the initial value
of $-2.7^\circ$ to $-6.0^\circ$, which is actually further
away from the final equilibrium value of $+2.7^\circ$. Thus,
the true magnitude of relaxation of tilt was ca.\ 8.7$^\circ$,
markedly larger than the above determined difference
of equilibrium values of 5.32$^\circ$. Since these initial 
oscillations cannot be fitted with any exponential-based 
function easily, the time $t=0$ was set to 164 fs for tilt,
where it reaches the global minimum; the fitted
function does not
describe the initial oscillations consequently. The results
of fitting are presented in Tables~\ref{tab:structrelax} and
\ref{tab:structrelaxtime}.

\begin{table*}
\begin{tabular}{lcccccccccccc}
\hline
              & $a_1$ & $t_1$ & $a_2$ & $t_2$ & $c_2$ & $a_4$ & $\tau$
              & RSS/TSS \\
\hline
Shift    & 0.27 & 0.281 & 1.28 & 55.5 & 0.568 & 1.89 & 603 & 0.00033 \\
uncert.       & 0.02 & 0.078 & 0.30 & 20.1 & 0.071 & 0.33 &  41 & \\
\hline
Tilt     & 1.48 & 0.133 & 3.28 & 61.4 & 0.633 & 3.53 & 685  & 0.0019 \\
uncert.       & 0.13 & 0.088 & 0.97 & 59.9 & 0.180 & 0.88 & 171  & \\
\hline
\end{tabular}
\caption{
Values of the parameters from the fitting of multi-exponential 
functions to the averaged data series of shift and tilt from the relaxation
simulations of electron transfer; $t_i$ in ps,
$a_i$ in \aa{}ngstr\"{o}m and degrees for shift and tilt,
respectively, $c_2$ dimensionless. The estimates of uncertainty
were obtained as described in the SI.}
\label{tab:structrelax}
\end{table*}

\begin{table*}
\begin{tabular}{lcccc}
\hline
 & \multicolumn{2}{c}{Shift}
 & \multicolumn{2}{c}{Tilt$^+$} \\
\hline
 & Magnitude & Decay Time
 & Magnitude & Decay Time \\
 & [\%] & [ps]
 & [\%] & [ps] \\
\hline
1 &  8 & 0.281    &  18 &  0.133    \\
2 & 37 & 90.1$^*$ &  40 &  86.5$^*$ \\
3 & 55 &   603    &  43 &   685     \\
\hline
\multicolumn{5}{l}{$^*$ stretched exponential} \\
\multicolumn{5}{p{4in}}{$^+$ To fit the tilt, $t=0$ is set 
at 164~fs, where the tilt reaches the global minimum due to 
the initial oscillation, see Figure S5. The relaxation is
considered to start at this point.} \\
\end{tabular}
\caption{
Decay times and relative magnitudes of the individual 
exponential components of the decays of shift and tilt.}
\label{tab:structrelaxtime}
\end{table*}

\begin{figure}[tb]
\includegraphics{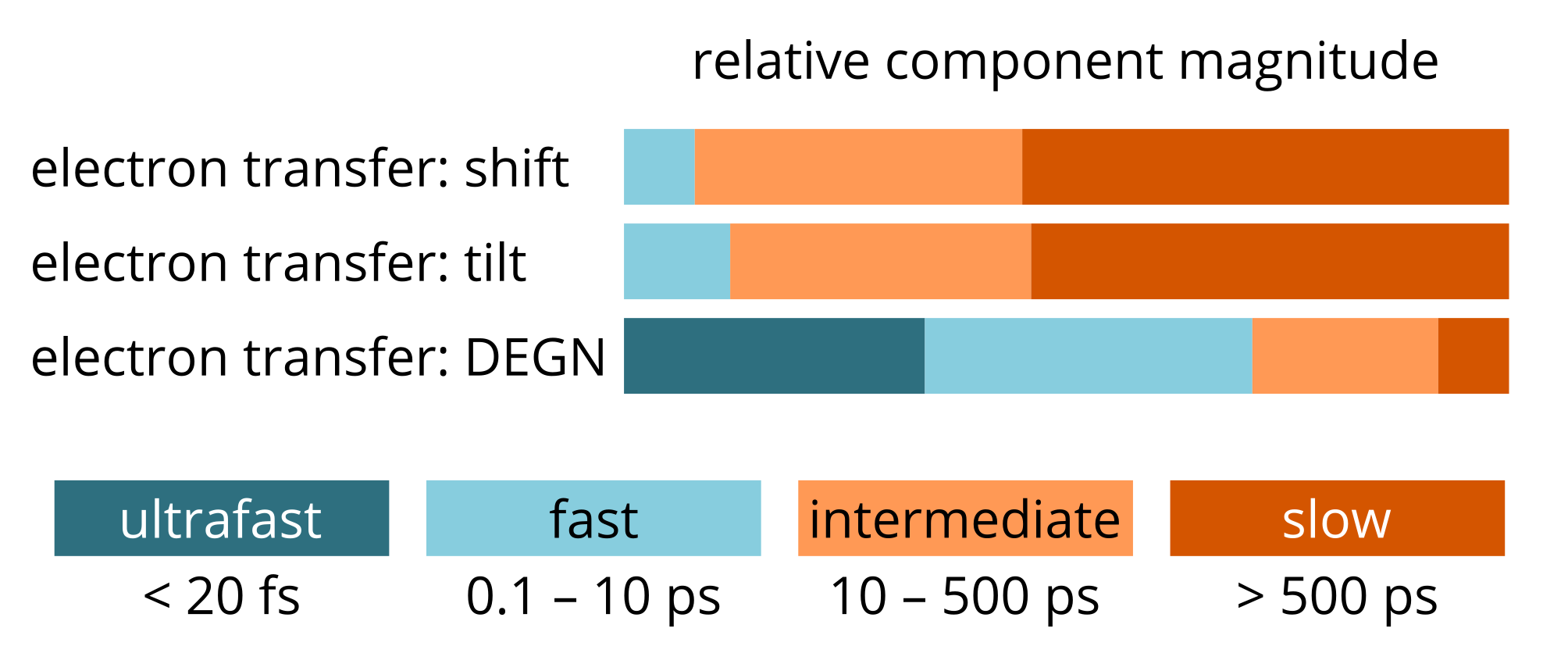}
\caption{Relative magnitudes of the various decay time
classes obtained for all of the relaxations investigated.}
\label{fig:times}
\end{figure}

The observed decay times are presented graphically in
Figure~\ref{fig:times}. When comparing the decays of helical
parameters, representing the structure of DNA, to the decay 
of DEGN, a striking difference is that there is no ultra-fast
decay component in the former. Recall that the relaxation of 
DEGN in Section~\ref{sec:relax-degn} contained a dominant 
component with a decay time in the femtosecond range and
magnitude of 34~\%. Apparently, there is no active
ultra-fast relative movement of nucleobases that would be 
responsible for this part of the DEGN decay. It may be 
concluded that rather than the DNA modes of motion, the dynamics
of solvent molecules are responsible, for instance the 
librational movements of water molecules. 

The structural relaxations are characterized by three kinds
of relaxations: i) a minor fast decay time of 0.1--0.3~ps,
ii) a major intermediate one of about 90~ps, and iii) a major
slow one of 600--700~ps. These observations are markedly 
different from the relaxation profiles of DEGN: The ultra-fast
component, extremely pronounced in DEGN, is missing altogether
in the structural decays. Further, the sub-picosecond 
relaxation has the lowest magnitude of the observed 
components. And finally, the slowest, nanosecond components
possess a much higher relative importance in the structural 
relaxation than in the DEGN relaxation.

Markedly, the slow component of the structural decay, responsible
for a quite large portion of the decay of both shift and tilt,
agrees with the slowest decay component of DEGN perfectly:
the decay times are 603 and 685~ps for shift and tilt, respectively,
while DEGN decays in 654~ps. This comparison suggests that the
longest relaxation time observed for DEGN (representing the 
interaction energies) is due to the dynamics of the DNA molecule
itself, as opposed to the shorter relaxation times, which may be
ascribed to the solvation dynamics. This scenario agrees with 
the interpretation of TRSS data collected for DNA by Sen \etal,
who stated that only 4~\% of the electric field correlations 
comes from the DNA \cite{Sen09}. In the current work, 8~\% of
DEGN is attributed to the DNA for the electron transfer.
A possible interpretation is that the closest solvent molecules
respond rapidly to electron transfer, accompanied
by hardly any change of the structure of the DNA itself. Only 
after a longer time, the DNA structure relaxes finally on two 
different time scales (90~ps and 600--700~ps), possibly 
together with an additional portion of the solvent.

Obviously, the different modes of motion of DNA and other 
components of the system, prominently, the solvent, contribute 
to the changes of energy with different strengths. There are even
some structural changes that do not contribute to the monitored 
energetics, which are simulation analogs of TRSS decays,
at all. Thus, no simple 
relationship can be deduced between the decay of energy and the
decay of structural parameters of DNA.

All of the decay times were determined on the basis of 
non-equilibrium simulations of a length of 1~ns, which 
turned out to be similar to the longest decay time.
Hence, it may be that any possible slower relaxation pattern is
hard to detect. To verify such possibility, additional extended 
simulations were performed, namely 100 simulations of 10~ns 
each for the relaxation following each of the electron transfer 
reactions $\mathrm{A}_4^+ \to \mathrm{A}_{10}^+$ and
$\mathrm{A}_{10}^+ \to \mathrm{A}_4^+$. All of the simulation
parameters were the same as in the simulations of 1~ns.
The resulting relaxation profiles are presented in Figure~S9.
No slower processes were observed, thus the lists of decay times
in Tables~\ref{tab:degnrelaxtime} and \ref{tab:structrelaxtime}
may be considered complete.


\subsection{Reaction Path Averaging}
\label{sec:results-rpa}

Results from the simulations described above have constituted 
a basis for an analysis of the structural relaxation of DNA upon 
electron transfer between two nucleobases, called here RPA,
which has proceeded in the following steps:
\begin{enumerate}
\item
A few coordinates of interest, or collective variables were 
defined. These are the base-pair step helical parameters of 
the central TA step, and shift and tilt were shown to be the 
most significant.
\item
These coordinates were recorded along a large number of 
realizations of an irreversible process, here, the relaxation 
of DNA structure described with classical MD simulation.
\item
The resulting time series of the collective variables were 
averaged. Because of the extremely strong variation of molecular
structures in the ensembles of irreversible simulations, the
above discussion of characteristic decay times was only made
possible by averaging of a huge number of MD trajectories 
generated.
\item
The averaged time series were analyzed further. The decay
times and magnitudes in Section~\ref{sec:relax-helical} were
obtained with fitting of multi-exponential functions to the
individual time series of collective variables. Additionally,
the time series were used to investigate correlations 
between the chosen collective variables, and to visualize
the studied irreversible process, specifically here, the 
averaged structural relaxation of DNA.
\end{enumerate}

\begin{figure*}[tb]
\includegraphics{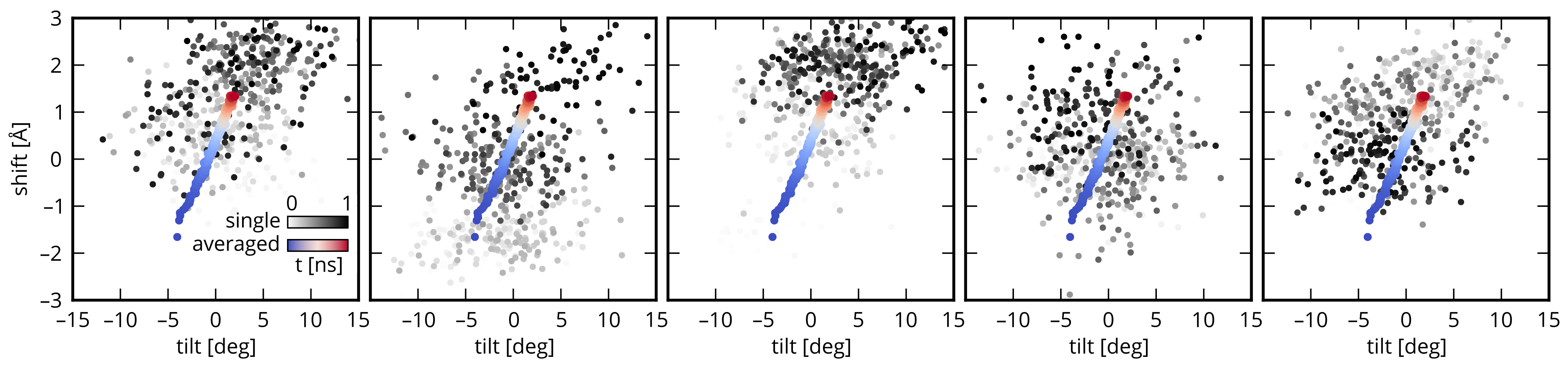}
\caption{The correlation of the base-pair step parameters shift 
and tilt following the electron transfer 
$\mathrm{A}_4^+\rightarrow\mathrm{A}_{10}^+$.
Time series from five individual trajectories are time-coded from white to black;
the averaged data are time-coded from blue to red.
Similar correlation plots for all of the other pairs of helical parameters
may be referred to in Figure~S7.}
\label{fig:structsingle}
\end{figure*}

The averaged time series of the shift and the tilt are shown
in correlation diagrams in Figure~\ref{fig:structsingle}, 
together with time series from a few examples of the many individual 
non-equilibrium trajectories. Two features are apparent immediately: 

The time series of shift and tilt in the individual non-equilibrium
trajectories are scattered all over the diagrams largely.
They are extremely different from each other,
thus they compose an overwhelmingly noisy ensemble
and do not reveal any correlation of shift and tilt whatsoever.
This vast structural variance is then removed by virtue of averaging,
which makes apparent the time course of the relaxation of shift and tilt,
and, in addition, their correlation.

Indeed, the averaged time series (blue-to-red in
Figure~\ref{fig:structsingle}) reveals a strong
linear correlation of shift and tilt, with a coefficient of
determination of $R^2=0.99$. This is identical to a previous
observation of the relaxation of shift and tilt proceeding on
the same time scales and with the same relative magnitudes
(Table~\ref{tab:structrelaxtime}), repeated from another 
perspective. The correlation of shift and tilt actually 
implies that the relaxation of DNA structure proceeds, for
the most part, along a single collective variable represented
by the linear combination of shift and tilt.
Correlations are also observed for other pairs of helical parameters,
see Figure~S7; interestingly, these are often non-linear, 
exhibiting more complex shapes. The same is true for the correlation
of relaxation profiles of DEGN with those of the helical 
parameters, see Figure~S8. Most importantly, any correlations
between the individual collective variables are only apparent 
after averaging of a large volume of non-equilibrium data.

The dynamics of the atomistic structure of the molecule
can be re-constructed from the averaged time series of the 
collective variables. Here, the time series of the helical 
parameters were translated into time-dependent coordinates 
of each atom, and were visualized, see the video clips in SI.

The first one (relaxation.mp4 in SI) captures the initial 
oscillations of the central base-pair step following
an instant transfer of a full electron between two adenines. 
The strong structural fluctuations dissipate within the first 
picosecond, being followed by a slower relaxation mostly in
the shift coordinate, compare with Figure~\ref{fig:structrelax}.
This movement proceeds in the direction of the final conformation,
as obtained from equilibrium simulations (shown in pale green).

The other video (transfer.mp4 in SI) shows structural distortions
of a longer dsDNA oligonucleotide accompanying a sequential electron
transfer along the molecule.
This visualization was constructed from DNA structures with helical
parameters obtained for the equilibrium states of the DNA molecule.
All of the long double helix except one base pair contained neutral
adenines, and the remaining one was considered as radical cation,
which was transferred along the DNA helix in a sequential manner.
For every elementary transfer event, the helical parameters were 
switched from the equilibrium values of the initial state to those
of the final state instantly.
Thus, the relaxation is considered instantaneous in this representation
in contrast to the detailed picture of a single base-pair step above.
The purpose of this video clip is to provide an idea of possible
change of structure of DNA during long-distance electron transfer.


\section{Discussion and Conclusions}
\label{sec:discuss}

We compare the time course of relaxation of DEGN as the probe
of energy relations in the system on the one hand, and the 
structural relaxation represented for the most part by the 
helical parameter shift, on the other. Since the relaxation 
is multi-modal and several individual ``energy modes'' are 
observed in DEGN, we investigate the possibility to identify 
each of the them with any of the modes of motion of the DNA.
In case no such mode is found, we would assume modes of motion
of the solvent to be responsible.
(Clearly, the current work cannot discriminate between the effects
of water and counterions as discussed previously.)\cite{Pal06a, Sen09}

The DEGN response observed in the current work is in accordance
with the previously reported components. First, there is 
a high-amplitude ultra-fast decay time of 8~fs, which
agrees well with the fastest component reported 
previously \cite{Pal06a}. Two further sizable contributions
of 1 ps and 10--12 or 20~ps were reported for de-excitation by 
Zewail \etal{}\cite{Pal03a,Pal03b}. A somewhat faster 
relaxation is observed following the electron transfer,
with decay times of 0.37 and 3.0~ps.
Third, there is a minor component with decay times of 
650~ps, which does not seem to correspond to any modes
of motion discussed previously in the context of 
hydration dynamics.

Three decay modes were found in the structural relaxations,
with decay times of 0.1--0.3~ps, 90~ps and 600--700~ps.
The first and second mode differ from the decay times of 
DEGN clearly, while the slowest one is especially interesting.
Although its relative importance in the DEGN relaxation is 
much lower than in the helical parameters, the remarkable 
similarity of the corresponding decay times for DEGN and 
helical parameters suggests that the longest relaxation time
observed for DEGN is actually due to the dynamics of the 
DNA molecule itself. This is opposed to the shorter relaxation
times, for which the solvation dynamics is most likely 
responsible; a similar conclusion was drawn previously by
Sen \etal{}\cite{Sen09}. Thus, the resulting picture of the
relaxation energetics, or the TRSS decay, is that the 
response of the closest solvent molecules drives its fastest
component(s), and the decay of the relaxation of DNA 
structure modulates the energies or TRSS on a longer 
time scale reaching a nanosecond.

Another important observation is more general: the different
modes of motion contribute to the changes of energy differently,
and some structural changes do not contribute to energies at all.
For instance, the fastest and second-fastest mode of decay of 
shift and tilt are clearly inactive in the DEGN decays.
Thus, there is no linear relationship between the decay of 
energy and the decay of structural parameters.

We also recall the meaning of DEG here. It is the difference of
potential energy in the system, obtained with the Hamiltonians 
corresponding to the initial and final state of the relaxation.
In our scheme, this reduces to the interaction energy of
the difference of atomic charges between the initial and
final state with the entire 
remainder of the molecular system (other nucleobases, DNA 
backbone and the solvent). Thus, effectively, this quantity
corresponds closely to the quantities used in the previous 
simulation studies, e.g.\ the difference of excited- and 
ground-state interaction energy \cite{Furse10} or the 
effective transition interaction energy \cite{Sen09}; by 
contrast, Pal \etal ~\cite{Pal06a, Pal06b} considered
simply the total ground-state interaction energy.

\subsection{Limitations of the Model}

Let us discuss the important approximations employed in our
methodology. It should be emphasized that all of them
apply to the previous simulation studies cited throughout this
work, as well.

In our simulations, perfectly equilibrated initial states
of the investigated processes were considered.
In reality, however, a radical cationic adenine may exhibit
a shorter lifetime; note the reported 
rate of electron transfer from one adenine to another of
4 or 20~ns$^{-1}$ in some DNA sequences \cite{Conron10,Takada04},
slightly faster than the slowest relaxation process.
Consequently, a full structural relaxation of DNA would take a longer
time to complete than an electron transfer.
Thus, the real structural distortions may be smaller in 
magnitude than the values in this work, which represent
the upper bounds of structural and energetic changes.

Another assumption is that the electron transfer reaction takes
place with exactly the same probability for every molecular structure 
in the equilibrium ensemble.
In reality, however, some configurations would be preferred,
mostly because of more appropriate activation energies, represented
for instance by the difference of instantaneous ionization 
energies close to zero.
This effect would be by far more time consuming to describe, while more
simplified approaches based on MM-only calculations of electric 
potentials are conceivable.
Difficulties may arise from the possibly inefficient sampling
of the configurations with high probability of electron transfer,
which may be rare events.

Also, the electron transfer process is considered to be
instantaneous, or infinitely fast,
and is modeled by a sudden and complete switch of MM atomic charges. 
It is unclear how electron transfer occurs in reality from the
microscopic point of view.
Note that this information does not seem to be accessible from
experiments, nor is covered by the classical theories of electron
transfer like Marcus' theory, and cannot be predicted by the current
non-adiabatic simulation schemes reliably \cite{Kubar13}.

Generally, the application of an empirical force field comes
with all of its advantages and drawbacks.
(i)
The recent parm99bsc0 force field used in this work describes the 
structure of DNA in a robust way, while some of its features
could still be improved \cite{Ivani16}.
(ii)
The derivation of atomic charges of the excited adenine molecule was
designed to resemble the recommended RESP methodology for ground state
as closely as possible, thus the adenine parameters are compatible 
with the remaining components of the force field.
(iii)
What classical MD simulation can never do is to describe the
dynamics of covalent bonds, mainly its quantum character.
This is one of the reasons why the MD simulations in this work were 
performed with all bond lengths constrained.
Consequently, any ultra-fast response of the DNA structure that
would be due to relaxation of bond lengths cannot be described.
Still, it is possible to consider the magnitude of energy relaxation
for which the change of bond lengths is responsible to be equal or smaller
than the inner-sphere reorganization energy for electron transfer.
The value for purine nucleobases is 0.23~eV,\cite{Kubar10}
which represents 6~\% of total energy relaxation,
thus, we consider this effect minor.
Also, we performed additional simulations of 4~$\mu$s of aggregated
sampling, having constrained only bonds involving hydrogen atoms.
The helical base-pair step parameters and their fluctuations,
see Table~S4, differ hardly from those obtained with all bonds constrained,
thus we consider the latter treatment justified.
(iv)
Another phenomenon impossible to describe is any interference of the 
dynamics of the nucleobase with the change of electron density
modeled by a set of fixed partial charges.

\subsection{Benefits}

Importantly, the methodology of the current study evinces
several advantageous points, novel in comparison with previous
works.

Most significantly, the true relaxation process is
modeled by performing non-equilibrium simulations. This is
in contrast to, and an improvement when comparing with most of the
cited previous work by other authors, who usually performed 
equilibrium simulations to describe non-equilibrium processes
within a linear response approximation. There is no need for
this transition in the current work.

The magnitudes of any structural changes are obtained
from extensive equilibrium MD simulations. Additionally, 
in this way, we provide the absolute values of baselines for 
the relaxation profiles conveniently, which are normally 
inaccessible via (auto)correlations. This is particularly
true about the structural parameters, whose changes can be
then converted to atomic models and visualized in 
a straightforward way.

In addition, the reaction path averaging concept was used to 
highlight the movements of DNA and visualized them in two 
videos clips. Itself, the idea of averaging non-equilibrium 
trajectories is not new. In 1990, Levy \etal{} investigated 
the relaxation of aqueous solvation shell of a formaldehyde 
molecule upon its de-excitation \cite{Levy90}. By means of 
averaging of 80 MD trajectories, it was possible
to identify the major mode of relaxation, which was the 
translational motion of water molecules. 
Laage and Hynes analyzed 16,000 hydrogen bond flipping events
observed in an equilibrium MD simulation of pure water, which made it
possible to propose a jump mechanism of water re-orientation \cite{Laage06}.
Some of the features of our RPA analysis resemble this 
previous work -- the introduction of suitable collective variables
as well as averaging of time series obtained for a large number of
realizations of the process of interest.
Most recently, the averaging of 10,000 non-equilibrium classical MD
simulations was used to infer energy dissipation in myoglobin
after CO photodissociation \cite{Brinkmann16}. In this work, the 
reaction was also modeled via instantaneous change of molecular 
mechanical force field.

Let us discuss the applicability of the RPA approach as 
implemented in this work to non-equilibrium chemical processes
in a more general context. There are a few requirements on the 
studied problem, and our application to processes taking place in
double-stranded DNA serves as an example of how they may be 
fulfilled.

Obviously, a suitable computational method for the description
of the process of interest has to be available. The current 
work made use of classical force field-based MD simulations,
which is perhaps the most convenient tool. Applications to 
non-equilibrium processes involving, e.g., extensive changes 
of electronic structure like re-arrangements of chemical 
bonding may require passing to quantum chemical or hybrid
QM/MM methodologies.

A sufficient number of non-equilibrium 
simulations has to be performed to yield converged results.
The number of simulations needed may easily become huge, 
placing excessive requirements on resources in terms of 
computational efficiency as well as storage space. Fortunately,
these calculations can be run in parallel trivially.

It seems to be favorable if the process of interest can be
described by means of (a small number of) collective variables
or reaction coordinates. This work benefited from the helical 
parameters for DNA being available. Also, the initial and final
states should be characterized first, for two reasons:
Fires, a structural ensemble of the initial state is needed to
provide initial conditions for the non-equilibrium simulations.
And second, it is a good idea to have estimated the extent of the
total change of the selected collective variables during the
process investigated. This information decreases the number of
fitting parameters for the time decays, and makes it possible 
to judge the convergence of the non-equilibrium simulations
to the ensemble of the final state.

\subsection{Summary and Outlook}

In summary, the structural changes in a double-stranded DNA 
oligonucleotide upon an electron transfer between two 
nucleobases were characterized. The dominant mode of motion 
is the helical parameter shift of the involved base-pair step.
The interaction energy decay following such a reaction 
were characterized as well. The contribution due to the
dynamics of the aqueous solvent seems to be dominant, while 
the relaxation of the DNA structure manifests itself on 
a longer time scale. The RPA approach provided a unique
illustration of DNA conformational changes, which would be 
otherwise difficult to visualize. Combined with a suitable
choice of collective variables, RPA may become an attractive
approach to characterize and visualize non-equilibrium
chemical processes in general.

Finally, there are a few directions in which this research 
may be extended.
For the specific application to DNA electron transfer,
the modes of motion of the surrounding water molecules are
of particular interest.
Therefore, a future RPA analysis may involve the solvation shell
of the nucleobases, described with appropriate coordinates,
or collective variables likely involving many water molecules.
An appropriate way to extend the RPA approach itself is to 
implement a weighting of the individual non-equilibrium trajectories
according to the propensity of the initial structures
to undergo the reaction.
Upon this modification, RPA will become
an even more realistic description of the non-equilibrium process.

\section*{Supporting Information}

Additional information on the simulation setup and the performed analyses;
detailed characterization of the equilibrium structure of the DNA molecule
involving the various states of the adenine nucleobase --
ground-state neutral, and radical cation;
additional graphical representation of the relaxation data;
results from alternative multi-exponential fits to relaxation data;
estimates of accuracy and uncertainty of fitting;
correlation diagrams of the DNA helical parameters in the course of relaxation;
two RPA videos depicting the relaxation of DNA structure upon electron transfer;
full author lists of Refs.~\citenum{Amber14}, \citenum{Olson01}, 
\citenum{Pasi14}, and \citenum{Ivani16}.


\begin{acknowledgement}

We thank Filip Lanka\v{s} for critical comments and valuable suggestions,
and we appreciate the encouragement by Pavel Hobza.
MHK is thankful for support provided by the Alexander von Humboldt 
Foundation. This research was also supported by the bwHPC
initiative and the bwHPC-C5 project, funded by the Ministry of 
Science, Research and the Arts Baden-W\"urttemberg (MWK) and the 
Research Foundation of Germany (DFG), through the associated compute
services of the JUSTUS HPC facility located at the University of Ulm.

\end{acknowledgement}


\clearpage

\section*{TOC Graphic}
\includegraphics{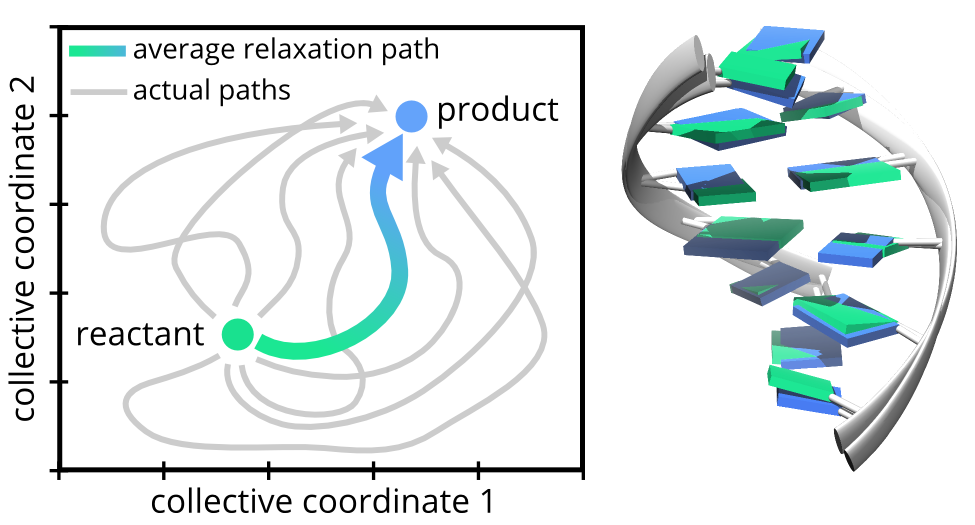}


\clearpage
\newpage
\onecolumn

\setcounter{equation}{0}
\setcounter{figure}{0}
\setcounter{table}{0}
\setcounter{page}{1}

\renewcommand\thefigure{S\arabic{figure}}
\renewcommand\thetable{S\arabic{table}}
\renewcommand\theequation{S\arabic{equation}}
\renewcommand\thepage{S\arabic{page}}
\renewcommand{\bibnumfmt}[1]{[S#1]}
\renewcommand{\citenumfont}[1]{S#1}
\newcommand{\std}{\mathrm{std}}
\newcommand{\sem}{\mathrm{sem}}

\begin{center}
\Huge{\bf Supporting Information to \\
``Reaction Path Averaging: \\
Characterizing Structural Response \\
of the DNA Double Helix \\
to Electron Transfer''}
\end{center}

\section{Methods}

\subsection{Simulation Box}

The atomic structure of a double-stranded DNA hexanucleotide with the
palindromic sequence d(CGTACG)$_2$ was prepared in the canonical B-DNA
conformation with the Nucleic Acid Builder. The DNA was 
placed in a periodic rhombic dodecahedron box. The size of the box was
chosen to ensure a distance of at least 1.2~nm between the DNA 
molecule and any box face. The box was filled with ca.\ 3000 water 
molecules, and 10 water molecules were replaced by sodium ions to 
compensate for the negative charge of DNA. The box contained 9427 
atoms in total.

The DNA was described with the Amber ff99bsc0 parameter
set \cite{SICornell95, SIPerez07}, translated into the Gromacs file format
with the ambconv utility \cite{SIRyjacek}. In order to avoid any 
possible fraying of the terminal base pairs, harmonic restraints
with a force constant of 2000 $\mathrm{kJ\,mol^{-1}\,nm^{-2}}$ were
applied on the heavy atoms of the three cytosine$\ldots$guanine 
hydrogen bonds of both terminal base pairs (i.e., 
CYT:N3$\ldots$GUA:N1, CYT:O2$\ldots$GUA:N2 and CYT:N4$\ldots$GUA:O6,
using the Amber naming convention). The three-site SPC/E water
model was used \cite{SIBerendsen87} because it yields reasonable
dynamic properties of bulk water as expressed in terms of 
self-diffusion coefficient and dielectric constant, much better
than e.g.\ the related TIP3P model does \cite{SIMark01}.
The sodium ions were modeled with parameters by Joung and 
Cheatham \cite{SIJoung08}.

\begin{figure}
\centering
\includegraphics{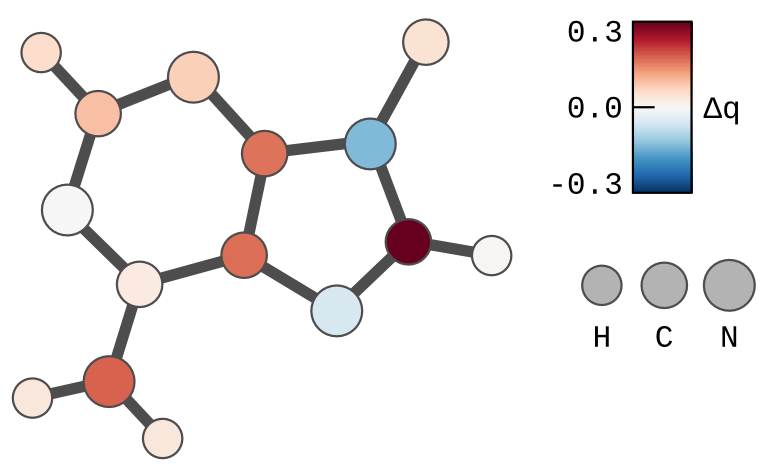}
\caption{The difference between partial atomic charge differences of the
charged and ground state (i.e. the default Amber ff99bsc0).}
\label{fig:charges}
\end{figure}

The adenine nucleobases were treated in a special way. The standard
Amber ff99bsc0 charges were used for the ground state, whereas 
modified charges were used for charged state 
(Figure~\ref{fig:charges}). Three distinct simulation setups were
employed, as
summarized in Table~\ref{tab:adenines}. Whenever one of 
the adenines was charged,
one of the sodium atoms was made neutral to maintain 
electroneutrality. Ten independent equilibrium simulations
differing in the initial velocities were run for each type of
simulation setup.

\begin{table}[bh]
\centering
\begin{tabular}{lrr}
\hline
\hline
abbrev &  A$_4$ charges &      A$_{10}$ charges \\
\hline
GS         &  ground state &      ground state \\
A$_4^+$    &  charged      &     ground state \\
A$_{10}^+$ &  ground state &      charged \\
\hline
\hline
\end{tabular}
\caption{Summary of simulation boxes.}
\label{tab:adenines}
\end{table}

\subsection{Simulation Setup}

The following set of equilibration simulations were performed
for each simulation setup:

\begin{itemize}

\item 1500 steps of conjugated gradient (CoG) energy minimization
using a steepest descent step every 10 CoG steps.

\item Assignment of random initial velocities from the 
Maxwell-Boltzmann distribution at 10~K.

\item $5\cdot 10^5$ steps of constant-volume water heating to
300~K using the v-rescale thermostat \cite{SIBussi07} with 
a coupling time of 0.2~ps. The solute was kept at 10~K with
an additional thermostat with the same coupling time.

\item $1\cdot 10^5$ steps of constant-volume heating to 300~K.
Two separate v-rescale thermostats with the coupling time of
0.5~ps were used for the solute and solvent.

\item $5\cdot 10^5$ steps of constant-pressure simulation 
at 300~K and 1~bar. Two separate v-rescale thermostats with
a coupling time of 0.5~ps were used together with a single
the Berendsen barostat \cite{SIBerendsen84} with a coupling 
time of 0.5~ps.

\item $2\cdot 10^6$ steps of constant-pressure simulation at
300~K and 1~bar using a single Nos\'{e}-Hoover 
thermostat \cite{SINose84, SIHoover85} ($\tau=0.5$~ps) and the
Parrinello-Rahman \cite{SIParrinello81} barostat ($\tau=0.5$~ps).

\item The electrostatics were treated with the Particle
Mesh Ewald algorithm \cite{SIDarden93} with a direct-space
cut-off of 1.1~nm and a maximum grid spacing of 0.12~nm.
The Lennard-Jones interactions were cut off at 1.1~nm.
A long-range correction for energy and pressure \cite{SIShirts07}
was applied. The equations of motion were integrated by means
of the leap-frog algorithm with a time step of 1~fs 
(NVT simulations) or 1.5~fs (NPT simulations). All bond
lengths were constrained with p-LINCS \cite{Hess08}.

\end{itemize}
For each simulation setup, ten independent trajectories
differing in the initial velocities were generated. The
production runs performed with the Nos\'{e}-Hoover thermostat
($\tau=1$~ps) and the Parrinello-Rahman barostat ($\tau=1$~ps) 
were extended to at least $3.4\cdot 10^8$ steps of 1.5~fs, 
corresponding to 510~ns. All of the other simulation
parameters were identical to the last equilibration 
step. A full-precision trajectory frame of the entire
simulation box was stored every 15~ps in order to initiate
non-equilibrium relaxations.

The non-equilibrium simulations were done at constant 
temperature of 300~K and pressure of 1~bar using identical
setup as in the equilibrium simulations. A time step of 
1~fs was used for time propagation, while all coordinates
were saved every 2~fs. The simulations were carried out 
with the Gromacs 4.6.7 program package \cite{SIPronk13, SIPall14}.

\subsection{Definitions of Statistical Descriptors}

\subsubsection{Mean Value}

The mean value $\left\langle A \right\rangle$ of
a quantity $A$ (e.g. helical parameter Shift),
measured for each frame $i$ of each of the trajectory $t$, was
calculated for the entire ensemble by equation $\ref{eq:meanmean}$.

\begin{equation}
\left \langle A \right \rangle = \frac{1}{N} \sum_t^N
\frac{1}{N_{f,t}} \sum_i^{N_{f,t}} A_i,
\label{eq:meanmean}
\end{equation}
where $N$ is the number of trajectories (i.e. 10), and $N_{f,t}$
is the number of frames for a given trajectory (typically about 350,000).
Each frame of each of the trajectories was taken with the same weight,
so the mean value could also be expressed by equation \ref{eq:mean}.

\begin{equation}
\left \langle A \right \rangle = \frac{1}{N_f} \sum_i^{N_f} A_i,
\label{eq:mean}
\end{equation}
where $N_f$ is the total number of frames in the ensemble.

\subsubsection{Standard Deviation}

The standard deviation $\std(A)$ of a quantity $A$ 
(e.g. helical parameter Shift), measured for each 
frame $i$, was calculated for the entire ensemble 
by equation $\ref{eq:std}$.

\begin{equation}
\std(A) = \sqrt{\frac{1}{N_f} \sum_i^{N_f} 
\left( A_i - \left \langle A \right \rangle \right)^2},
\label{eq:std}
\end{equation}
where $N_f$ is the total number of frames 
and $\left \langle A \right \rangle$
stands for the mean value (Equation \ref{eq:mean}).

\subsubsection{Standard Error of the Mean}

The standard error of the mean $\sem(A)$ of a quantity $A$
(e.g., helical parameter Shift), measured for each frame $i$,
was calculated for the entire ensemble by equation
$\ref{eq:sem}$.

\begin{equation}
\sem(A) = \frac{\std(A)}{\sqrt{N_f}},
\label{eq:sem}
\end{equation}
where $N_f$ is the total number of frames and $\std(A)$ stands for
the standard deviation as calculated by equation \ref{eq:std}.

\subsubsection{Difference of Mean Values}

The difference $\Delta A$ of two quantities $A_1$ and $A_2$ 
(e.g. helical parameter
Shift of the excited-state and ground-state ensembles, 
$\Delta$Shift) was calculated as a difference of the mean values
of the final $\left\langle A_2 \right\rangle$ and 
initial $\left\langle A_1 \right\rangle$ states 
(equation \ref{eq:diff}).

\begin{equation}
\Delta A = \left \langle A_2 \right \rangle -
\left \langle A_1 \right \rangle.
\label{eq:diff}
\end{equation}

\subsubsection{Standard deviation of the Difference}

The standard deviation $std(\Delta A)$ of 
a difference of two quantities $A_1$ and $A_2$ 
(e.g. helical parameter
Shift of the excited-state and ground-state ensembles, 
$\Delta$Shift) was calculated from the standard deviations
of the two quantities over all ensembles by
equation \ref{eq:stddiff}.

\begin{equation}
\std(\Delta A) = \sqrt{\std(A_1)^2 + \std(A_2)^2}
\label{eq:stddiff}
\end{equation}

\subsubsection{Standard Error of the Mean Difference}

The standard error of the mean of a difference $\sem(\Delta A)$
of two quantities $A_1$ and $A_2$ (e.g., helical parameter
Shift of the excited-state and ground-state ensembles, 
$\Delta$Shift) was calculated as a sum of standard errors 
of the mean of each of the quantities following equation
\ref{eq:semdiff}.

\begin{equation}
\sem(\Delta A) = \sem(A_1) + \sem(A_2)
\label{eq:semdiff}
\end{equation}

\clearpage
\newpage

\section{Results and Discussion}

\subsection{Schematic Representations of Helical Parameters}

\begin{figure}[hb]
\centering
\includegraphics{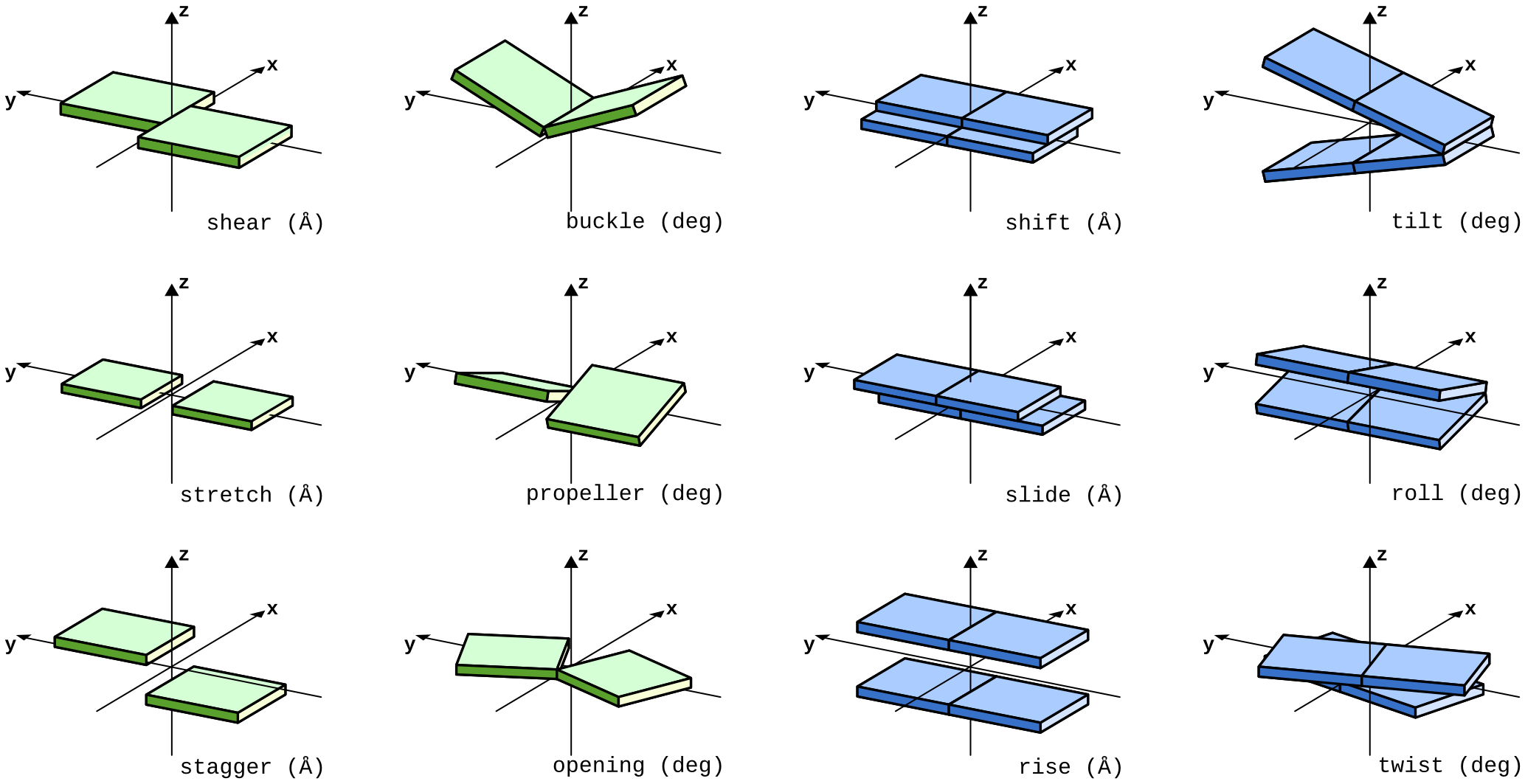}
\caption{The collective coordinates for the description of DNA
structure (a.k.a.\ helical parameters) \cite{SIDickerson89} 
in the standard reference frame as defined by Olson 
\etal \cite{SIOlson01}. Each box represents a nucleobase.
The parameters of base pairs are in green, and those of
base-pair steps are blue.}
\label{fig:helparams}
\end{figure}

\clearpage
\newpage

\subsection{Equilibrium Simulations}

\subsubsection{Helical Parameters}

\begin{table}[hb]
\centering
\begin{tabular}{lrcrc}
\hline
\hline
~ &
\multicolumn{2}{c}{T$_3$-A$_{10}$ pair} &
\multicolumn{2}{c}{A$_4$-T$_9$ pair} \\
\hline
~ & 
$\mathrm{mean}\pm\mathrm{std}$ & $\mathrm{sem} \times 10^3$ &
$\mathrm{mean}\pm\mathrm{std}$ & $\mathrm{sem} \times 10^3$ \\
\hline
Shear     & $-0.04 \pm0.37 $& 0.2 & $  0.05\pm 0.34$& 0.2 \\
Stretch   & $ 0.00 \pm0.14 $& 0.1 & $  0.00\pm 0.14$& 0.1 \\
Stagger   & $ 0.15 \pm0.42 $& 0.2 & $  0.15\pm 0.42$& 0.2 \\
Buckle    & $-1.65 \pm11.49$& 6.0 & $  1.45\pm11.50$& 6.0 \\
Propeller & $-12.01\pm8.60 $& 4.5 & $-12.01\pm 8.64$& 4.5 \\
Opening   & $-0.08 \pm6.20 $& 3.2 & $ -0.05\pm 6.24$& 3.3 \\
\hline
\hline
\end{tabular}
\caption{Mean values (mean), standard deviations (std) and standard
errors of the mean (sem) of the helical parameters of the two central TA
base pairs calculated from the ground-state equilibrium trajectories.}
\end{table}

\begin{table}[hb]
\centering
\begin{tabular}{lrccc}
\hline
\hline
~ &
\multicolumn{2}{c}{simulations} &
A-DNA$^a$ & B-DNA$^a$ \\
\hline
~ & $\mathrm{mean}\pm\mathrm{std}$ & $\mathrm{sem} \times 10^3$ & 
$\mathrm{mean}\pm\mathrm{std}$ & $\mathrm{mean}\pm\mathrm{std}$ \\
\hline
Shift$^b$ & $  0.01\pm0.89$ & 0.5 & $0.00\pm0.54$  & $-0.02\pm0.45$ \\
Slide 	  & $ -0.21\pm0.77$ & 0.4 &$-1.53\pm0.34$ & $0.23\pm0.81$ \\
Rise      & $  3.21\pm0.32$ & 0.2 & $3.32\pm0.20$  & $3.32\pm0.19$ \\
Tilt$^b$  & $ -0.03\pm4.97$ & 2.6 & $0.1\pm2.8$   & $0.1\pm2.5$ \\
Roll      & $  8.01\pm7.28$ & 3.8 & $8.0\pm3.9$   & $0.6\pm5.2$ \\
Twist     & $ 32.34\pm6.60$ & 3.4 & $31.1\pm3.7$  & $36.0\pm6.8$ \\
\hline
\hline
\multicolumn{5}{l}{
$^a$ sequence-averaged data from X-ray crystallography \cite{Olson01}
} \\
\multicolumn{5}{l}{$^b$ should be zero due to helix anti-symmetry }\\
\end{tabular}
\caption{Mean values (mean), standard deviations (std) and standard
errors of the mean (sem) of the helical parameters of the central TA
base-pair step calculated from the ground-state equilibrium trajectories.}
\label{tab:gsstep}
\end{table}

\begin{table}[hb]
\centering
\begin{tabular}{lrcrcrc}
\hline
\hline
 &
\multicolumn{2}{c}{$\mathrm{A}_4^+$} &
\multicolumn{2}{c}{$\mathrm{A}_{10}^+$} &
\multicolumn{2}{c}{$\mathrm{A}_4^+$ (flexi) } \\
\hline
 &
$\mathrm{mean}\pm\mathrm{std}$ & $\mathrm{sem} \times 10^3$ &
$\mathrm{mean}\pm\mathrm{std}$ & $\mathrm{sem} \times 10^3$ &
$\mathrm{mean}\pm\mathrm{std}$ & $\mathrm{sem} \times 10^3$ \\
\hline
Shift & $ -1.70\pm0.89$ & 0.5   & $  1.70\pm0.88$ & 0.4 & $-1.80\pm0.84$ & 0.5 \\
Slide & $  0.31\pm0.52$ & 0.3   & $  0.33\pm0.52$ & 0.3 & $ 0.37\pm0.48$ & 0.3 \\
Rise  & $  3.12\pm0.30$ & 0.2   & $  3.13\pm0.29$ & 0.1 & $ 3.14\pm0.30$ & 0.2 \\
Tilt  & $ -2.64\pm5.07$ & 2.6   & $  2.68\pm5.06$ & 2.5 & $-3.04\pm5.00$ & 3.1 \\
Roll  & $ -0.56\pm6.90$ & 3.5   & $ -0.67\pm6.87$ & 3.4 & $-0.83\pm6.84$ & 4.2 \\
Twist & $ 37.39\pm5.99$ & 3.1   & $ 37.85\pm5.79$ & 2.9 & $37.84\pm6.20$ & 3.8 \\ 
\hline
\hline
\end{tabular}
\caption{Mean values (mean), standard deviations (std) and standard
errors of the mean (sem) of the helical parameters of the central TA
base-pair step calculated from the equilibrium trajectories with 
radical-cation adenine. The last two columns (A$_4^+$ flexi) represent the 
trajectories obtained with flexible bonds between heavy atoms.}
\label{tab:qstep}
\end{table}

\clearpage
\newpage

\subsubsection{Sugar Puckering and Backbone Substates}

\begin{table}[hb]
\centering
\begin{tabular}{lcccc}
\hline
\hline
~ & T$_3$ & A$_{10}$ & A$_4$ & T$_9$ \\
GS &
417.2$\pm$926.5 & 466.3$\pm$1012.6 & 439.4$\pm$922.0 & 387.5$\pm$873.1 \\
A$_4^+$ &
16.4$\pm$53.6 & 158.6$\pm$432.8 & 3552.3$\pm$6210.3 & 1279.2$\pm$3136.0 \\
A$_{10}^+$ &
1288.4$\pm$3283.4 & 3585.9$\pm$5961.1 & 171.0$\pm$492.7 & 16.3$\pm$50.7 \\
\hline
\hline
\label{tab:b1times}
\end{tabular}
\caption{Life times of the BI backbone substate in the form of mean value
$\pm$ standard deviation. Derived from the equilibrium data.}
\end{table}

\begin{table}[hb]
\centering
\begin{tabular}{lcccc}
\hline
\hline
~ & T$_3$ & A$_{10}$ & A$_4$ & T$_9$ \\
GS &
18.7$\pm$63.6  &  73.7$\pm$243.9 &  75.7$\pm$233.3 & 17.5$\pm$66.0 \\
A$_{4}^+$ &
35.5$\pm$94.2  & 206.2$\pm$530.9 &  38.5$\pm$144.1 & 46.4$\pm$144.7 \\
A$_{10}^+$ &
47.8$\pm$142.2 &  33.6$\pm$118.0 & 189.5$\pm$486.5 & 36.1$\pm$77.7 \\
\hline
\hline
\label{tab:b2times}
\end{tabular}
\caption{Life times of the BII backbone substate in the form of mean value
$\pm$ standard deviation. Derived from the equilibrium data.}
\end{table}

\begin{table}[hb]
\centering
\begin{tabular}{lrrrr}
\hline
\hline
     pucker &     T3 &     A10 &      A4 &      T9 \\
\hline
\multicolumn{5}{l}{A$_{4}^+$} \\
   C1'-exo &    52 &    48 &     8 &    52 \\
  C2'-endo &    31 &    31 &    54 &    36 \\
   C3'-exo &     0 &     4 &    35 &     1 \\
   C4'-exo &     7 &     4 &     1 &     2 \\
  O4'-endo &    10 &    12 &     2 &    10 \\
\hline
\multicolumn{5}{l}{A$_{10}^+$} \\
  C1'-exo  &    51 &     7 &    48 &    52 \\
  C2'-endo &    36 &    54 &    32 &    31 \\
   C3'-exo &     2 &    35 &     4 &     0 \\
   C4'-exo &     2 &     1 &     4 &     6 \\
  O4'-endo &    10 &     2 &    12 &    10 \\
\hline
\hline
\end{tabular}
\caption{Populations (in \%) of the different deoxyribose ring
conformations (i.e., sugar pucker) of the nucleobases in the 
central base pair step in charged states.}
\label{tab:puckers}
\end{table}

\clearpage
\newpage

\subsubsection{End-State Changes of Helical Parameters}

\begin{table}[hb]
\centering
\footnotesize
\begin{tabular}{lrcrc}
\hline
\hline
~ &
\multicolumn{2}{c}{T$_3$-A$_{10}$} &
\multicolumn{2}{c}{A$_4$-T$_{9}$} \\
\hline
~ & 
mean$\pm$std & $\mathrm{sem} \times 10^3$ &
mean$\pm$std & $\mathrm{sem} \times 10^3$ \\
\hline
$\Delta$Shear    &  $0.27\pm 0.50$ & 0.4 &  $0.24\pm 0.64$ & 0.5  \\
$\Delta$Stretch  & $-0.06\pm 0.21$ & 0.1 & $-0.06\pm 0.21$ & 0.1  \\
$\Delta$Stagger  &  $0.15\pm 0.57$ & 0.4 &  $0.16\pm 0.57$ & 0.4  \\
$\Delta$Buckle   &  $0.88\pm15.47$ & 11  &  $1.19\pm15.45$ & 11   \\
$\Delta$Propeller&  $7.88\pm12.20$ & 8.7 & $-7.23\pm12.06$ & 8.6  \\
$\Delta$Opening  &  $5.70\pm 8.36$ & 5.9 & $-5.35\pm 8.39$ & 5.9  \\
\hline
\hline
\end{tabular}
\caption{Overall change in helical base-pair parameters of the
third T$_3$-A$_{10}$ and fourth A$_4$-T$_9$ base pairs upon
$\mathrm{A}_{10}^+\rightarrow\mathrm{A}_4^+$ reaction
as represented by differences of mean values (mean) calculated from
the equilibrium ensembles equivalent to at least
5.1~$\mu$s each. Standard deviations (std) and standard errors of the
mean (sem) are also provided. Shear, stretch and stagger in \AA; buckle,
propeller and opening in degrees. Note that 
the $\mathrm{A}_{10}^+\rightarrow\mathrm{A}_4^+$ reaction yields
negative values of $\mathrm{A}_{4}^+\rightarrow\mathrm{A}_{10}^+$ 
by definition.}
\label{tab:bpparams3}
\end{table}

\clearpage
\newpage

\subsection{Equilibrium Probability Density Functions}

\begin{figure}[hb]
\centering
\includegraphics{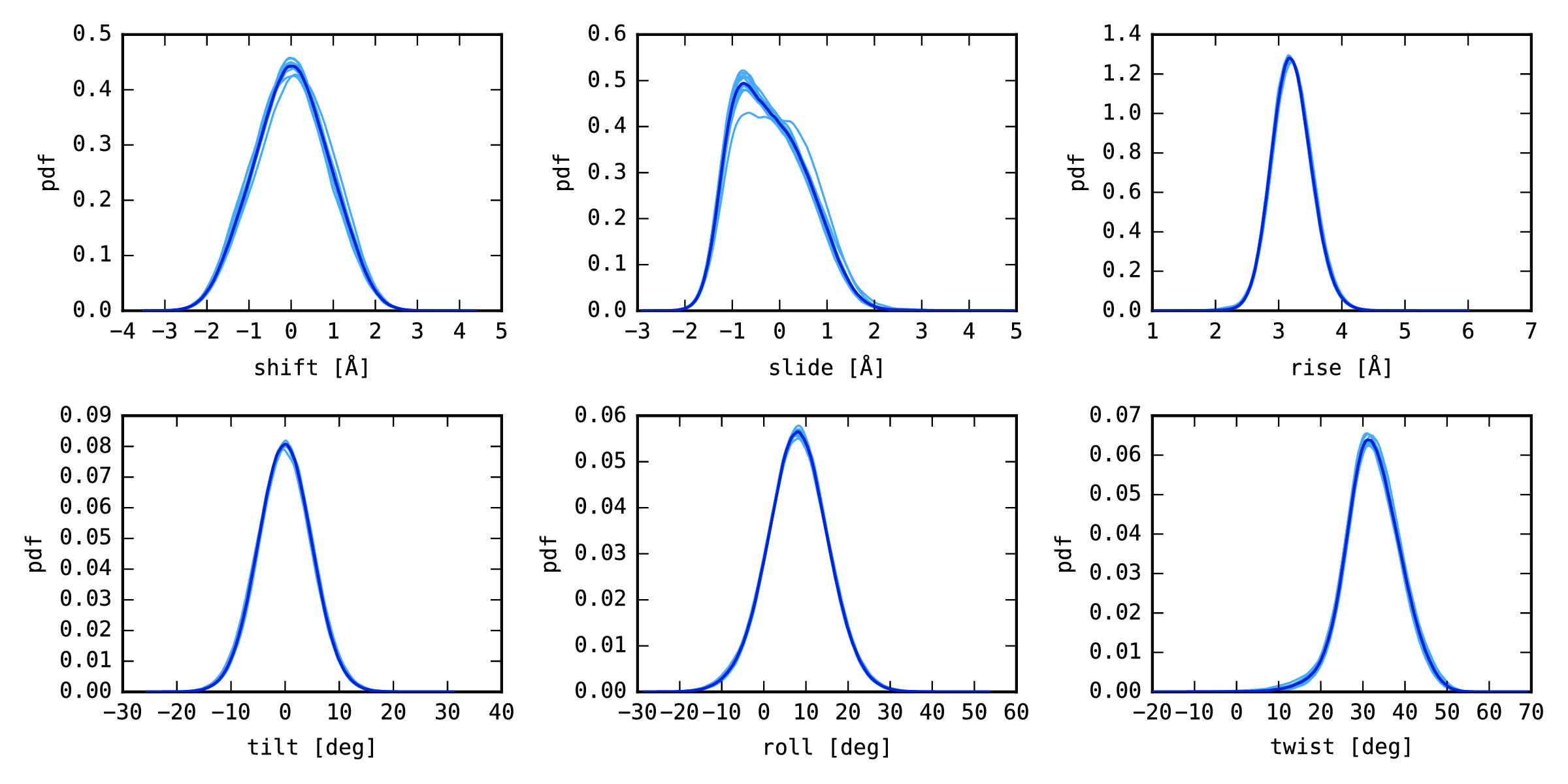}
\caption{Probability density functions (pdf) of the parameters 
of the central base-pair step TA, shown for ten 510-ns-long 
trajectories (light blue), and for the aggregated ensemble
(dark blue), where both adenines were in the ground electronic
state.}
\label{fig:kdegs}
\end{figure}

\begin{figure}[hb]
\centering
\includegraphics{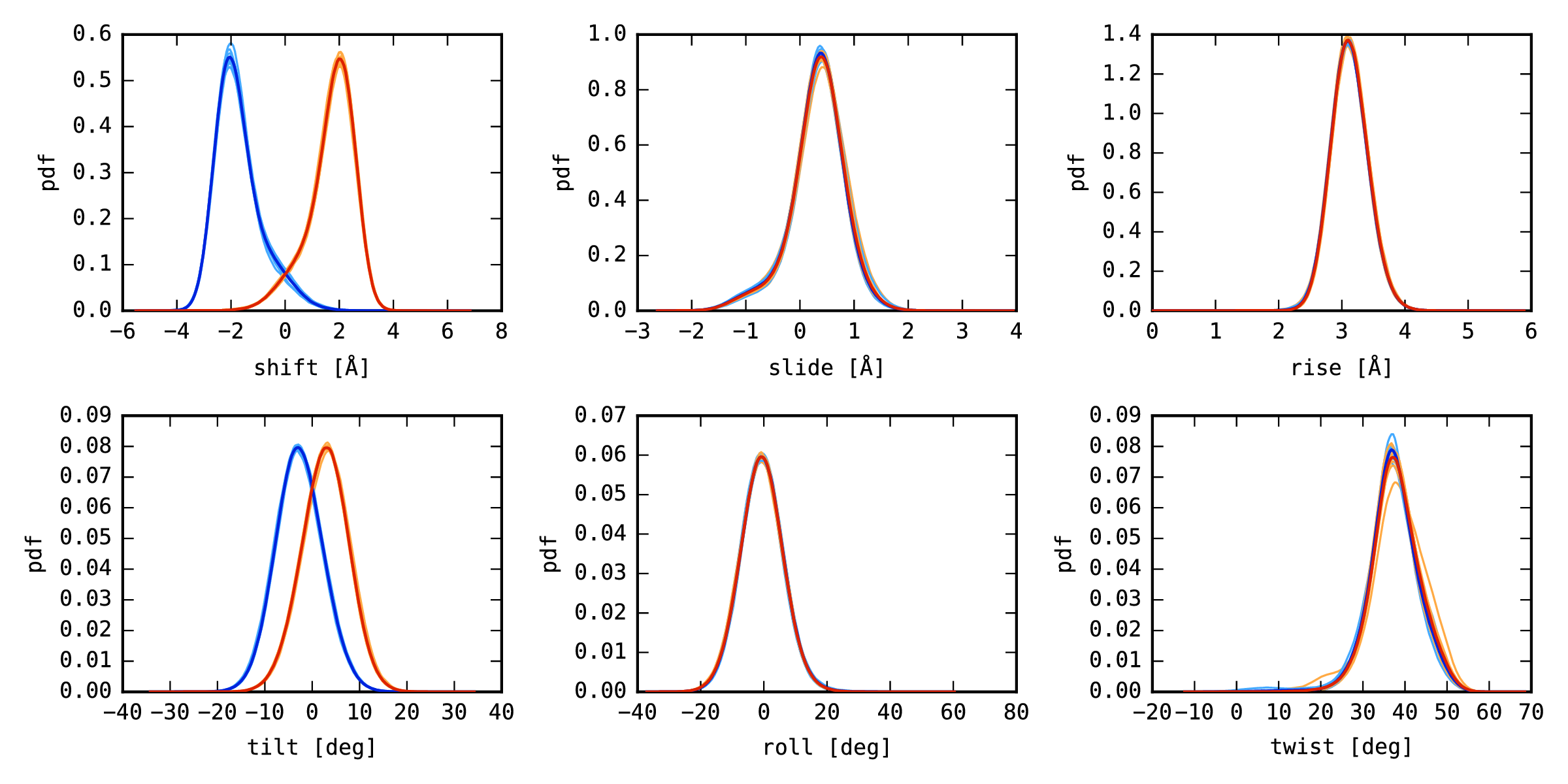}
\caption{Probability density functions (pdf) of the parameters 
of the central base-pair step TA, shown for ten 510-ns-long 
trajectories (light blue), and for the aggregated ensemble
(dark blue), where either A$_4$ (blue) or A$_{10}$ (orange) 
were in the cationic form.}
\label{fig:kdeq}
\end{figure}

\clearpage
\newpage

\subsection{Fitting Relaxation Data}

The results of alternative fitting choices on DEGN times series are shown
in Table~\ref{tab:fit-degn} together with the final ones.
Recall the fitting function:
\B f(t) = a_1 \exp \left[ -\frac{t}{t_1} \right]
        + a_2 \exp \left[ -\left( \frac{t}{t_2} \right)^{\!\! c_2} \right]
        + a_3 \exp \left[ -\left( \frac{t}{t_3} \right)^{\!\! c_3} \right]
        + a_4 \exp \left[ -\frac{t}{\tau} \right]
\label{eq:fit-degn} \E
A single exponential and a single stretched-exponential fit are shown
to demonstrate that they are insufficient,
and also to provide an idea of the typical magnitudes of RSS/TSS.
The extremely stretched exponential ($c_2 = 0.146$)
performs perhaps better than expected.

\begin{table}[h]
\footnotesize
\centering
\begin{tabular}{lcccccccccccccc}
\hline
\hline
DEGN      & $a_1$ & $t_1$ & $a_2$ & $t_2$ & $c_2$ &
            $a_3$ & $t_3$ & $c_3$ & $a_4$ & $\tau$ & RSS/TSS \\
\hline
Q         & 0.34 & 0.0080 & 0.37 & 0.198 & 0.516 &
            0.21 & 1.43 & 0.488 & 0.08 & 654       & 0.0076 \\
uncert.   & 0.03 & 0.0004 & 0.11 & 0.052 & 0.090 &
            0.08 & 1.62 & 0.104 & 0.01 & 95        \\
\hline
Q         & 0.23 & 0.0090 & 0.69 & 0.240 & 0.367 &
                 &        &     & 0.08 & 654      & 0.0082 \\
uncert.   & 0.05 & 0.0004 & 0.06 & 0.051 & 0.032 &
                 &        &     & 0.01 & 91        \\
\hline
Q         &      &        & 1    & 0.431 &       &
                 &        &     &      &          & 2.74 \\
\hline
Q         &      &        & 1    & 0.040 & 0.146 &
                 &        &     &      &          & 0.110 \\
\hline
\hline
\end{tabular}
\caption{Parameters from the fitting of Eq.~\ref{eq:fit-degn} to the data series of DEGN;
$t_i$ in ps.}
\label{tab:fit-degn}
\end{table}

The results of alternative fitting choices on the times series of shift and tilt
are shown in Table~\ref{tab:fit-helical} together with the final ones.
A four-exponential function was tested also:
\B f(t) = a_1 \exp \left[ -\frac{t}{t_1} \right]
          + a_2 \exp \left[ -\left( \frac{t}{t_2} \right)^{\!\! c_2} \right]
          + a_3 \exp \left[ -\frac{t}{t_3} \right]
          + a_4 \exp \left[ -\frac{t}{\tau} \right]
\label{eq:fit-helical} \E

\begin{table}[ht]
\footnotesize
\centering
\begin{tabular}{lcccccccccccccc}
              & $a_1$ & $t_1$ & $a_2$ & $t_2$ & $c_2$
              & $a_3$ & $t_3$ & $a_4$ & $\tau$ & RSS/TSS \\
\hline
\hline
Q shift & 0.27 & 0.281 & 1.28 & 55.5 & 0.568
        &      &       & 1.89 & 603 & 0.00033 \\
uncert. & 0.02 & 0.078 & 0.30 & 20.1 & 0.071
              &      &       & 0.33 &  41 & \\
\hline
Q shift & 0.30 & 0.272 & 0.48 & 9.47 & 0.724
        & 0.70 &  107  & 1.96 & 603 & 0.00027 \\
uncert. & 0.05 & 0.040 & 0.20 & 5.81 & 0.365
        & 0.26 &   52  & 0.27 &  46 & \\
\hline
Q shift & 0.37 &  4.15 & 0.83 & 85.3 &       
        &      &       & 1.98 & 603 & 0.00073 \\
\hline
\hline
Q tilt  & 1.48 & 0.133 & 3.28 & 61.4 & 0.633 &      &      & 3.53 & 685  & 0.0019 \\
uncert. & 0.13 & 0.088 & 0.97 & 59.9 & 0.180 &      &      & 0.88 & 171  & \\
\hline
Q tilt  & 1.55 & 0.0896& 1.15 & 6.66 & 0.629 & 2.15 & 99.6 & 3.65 & 685  & 0.0018 \\
uncert. & 0.23 & 0.0639& 0.35 & 3.75 & 0.332 & 0.62 & 48.4 & 0.56 & 127  & \\
\hline
Q tilt  & 0.86 & 3.59  & 2.42 & 86.3 &       &      &      & 3.68 & 685  & 0.0030 \\
\hline
\hline
\end{tabular}
\caption{Parameters from the fitting of Eq.~\ref{eq:fit-helical}
to the data series of shift and tilt.
$t_i$ in ps;
$a_i$ in \aa{}ngstr\o{}m and degrees for shift and tilt, respectively.}
\label{tab:fit-helical}
\end{table}

Among the data series from electron transfer simulations,
the fits of the shift and the tilt behave similarly, and are marginally improved
upon the inclusion of a fourth exponential.
On the other hand, considering the component $a_2$ non-stretched
deteriorates the fit considerably.

\begin{figure}[b]
\centering
\includegraphics[scale=0.9]{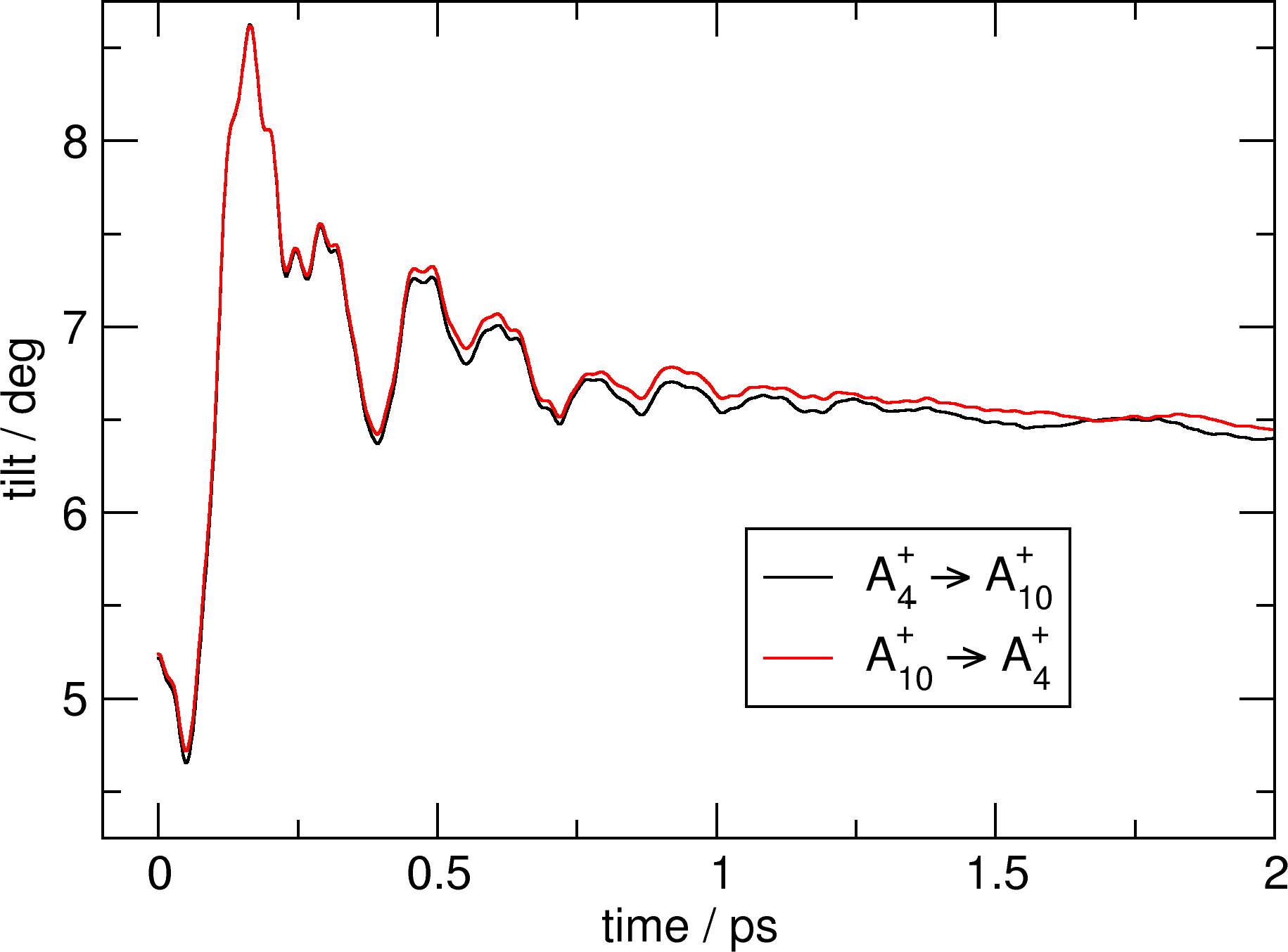}
\caption{Relaxation of the tilt in the first 2~ps of the electron transfer
simulations.}
\label{fig:tilt-start}
\end{figure}

\clearpage
\newpage

\subsubsection{Fitting accuracy and uncertainty}

The accuracy of the obtained fits was judged by the ratio
of the residual sum of squares and the total sum of 
squares, RSS/TSS, given as
\B
\textrm{RSS/TSS} = \frac{
\sum_i^n \left( y_i - f(x_i) \right)^2
}{
\sum_i^n \left( y_i - \bar{y} \right)^2
}
\E
for a data series $(x_i,y_i)$ ($i=1,\ldots{},n$) with
a mean value of $\bar{y} = \sum_i^n y_i / n$, and a fitted
function $f(x)$.

The statistical uncertainty of the parameters was estimated as follows:
The entire ensemble of data series was divided into ten disjunctive groups,
so that each group contained 10,000 shorter data series and 100 longer
data series.
Then, the data series in each of the groups were averaged to produce
a single data series representative of the respective group,
which was subject to fitting with the function
in Eq.~\ref{eq:fit-degn} or \ref{eq:fit-helical}.
Thus, ten values were obtained for each of the parameters of the fitting function.
Finally, standard deviation was calculated for each of the parameters,
from the values obtained from fitting to the `small' data series.
These values quantify the uncertainty of determination of that parameter.
The line `tot' contains the parameters obtained from fitting to the entire data series,
while each numbered line contains values from a fit to a data series obtained
by averaging over one tenth of the available data.
The result is a sample standard deviation
$\sigma (x) = \sqrt{\sum_i (x_i - \bar{x})^2 / (n-1)}$
obtained for the ensemble of the respective parameter $x$ in the table,
excluding any outlier fits, marked `*';
$\bar{x}$ is the average value of the parameter $x$,
and $n$ is the number of non-outlier values considered.

\begin{table}[hb]
\centering
\begin{tabular}{lccccccccccc}
fit & RSS/TSS & $a_1$ & $t_1$ & $a_2$ & $t_2$ & $c_2$ 
              & $a_3$ & $t_3$ & $c_3$ & $a_4$ & $\tau$  \\
\hline
0  & 0.0109 & 0.38 & 0.0073 & 0.23 & 0.129 & 0.714 & 0.30 & 1.08 & 0.58 & 0.092 & 517 \\
1  & 0.0114 & 0.33 & 0.0082 & 0.49 & 0.267 & 0.471 & 0.11 & 3.68 & 0.43 & 0.073 & 710 \\
2  & 0.0111 & 0.34 & 0.0080 & 0.38 & 0.205 & 0.499 & 0.19 & 1.45 & 0.47 & 0.086 & 537 \\
3* & 0.0129 & 0.38 & 0.0080 & 0.07 & 0.124 & 2.683 & 0.45 & 0.58 & 0.49 & 0.092 & 613 \\
4  & 0.0121 & 0.34 & 0.0080 & 0.41 & 0.232 & 0.502 & 0.17 & 1.83 & 0.45 & 0.080 & 691 \\
5  & 0.0115 & 0.29 & 0.0086 & 0.57 & 0.281 & 0.429 & 0.06 & 5.89 & 0.33 & 0.071 & 643 \\
6* & 0.0158 & 0.36 & 0.0079 & 0.07 & 0.124 & 2.844 & 0.48 & 0.52 & 0.47 & 0.090 & 753 \\
7  & 0.0105 & 0.34 & 0.0081 & 0.39 & 0.210 & 0.511 & 0.18 & 1.71 & 0.50 & 0.087 & 615 \\
8  & 0.0128 & 0.34 & 0.0081 & 0.45 & 0.232 & 0.498 & 0.13 & 2.84 & 0.54 & 0.079 & 790 \\
9  & 0.0129 & 0.30 & 0.0086 & 0.56 & 0.293 & 0.431 & 0.08 & 3.85 & 0.27 & 0.068 & 732 \\
\hline
tot& 0.0076 & 0.34 & 0.0080 & 0.37 & 0.198 & 0.516 & 0.21 & 1.43 & 0.49 & 0.083 & 654 \\
$\sigma$ &  & 0.03 & 0.0004 & 0.11 & 0.052 & 0.090 & 0.08 & 1.62 & 0.10 & 0.008 &  95 \\
\hline
\end{tabular}
\caption{Electron transfer -- DEGN.}
\end{table}

\begin{table}[hb]
\centering
\begin{tabular}{lccccccccccc}
fit & RSS/TSS & $a_1$ & $t_1$ & $a_2$ & $t_2$ & $c_2$ 
              & $a_4$ & $\tau$  \\
\hline
0  & 0.00208 & 0.29 & 0.322 &   1.42 &  61.2 & 0.652 &    1.64 & 637 \\ 
1* & 0.00326 & 0.28 & 0.300 &   2.83 & 234.8 & 0.582 &    0.34 & 580 \\ 
2  & 0.00171 & 0.30 & 0.361 &   1.50 &  74.2 & 0.640 &    1.69 & 553 \\ 
3  & 0.00391 & 0.31 & 0.339 &   0.91 &  32.3 & 0.679 &    2.23 & 645 \\ 
4  & 0.00344 & 0.32 & 0.375 &   1.09 &  40.0 & 0.712 &    2.00 & 677 \\ 
5* & 0.00296 & 0.35 & 3.652 & 262.58 & 529.8 & 0.998 &--259.99 & 531 \\ 
6  & 0.00205 & 0.30 & 0.276 &   0.70 &  15.8 & 0.727 &    2.56 & 601 \\ 
7  & 0.00207 & 0.32 & 0.378 &   0.79 &  27.7 & 0.775 &    2.30 & 583 \\ 
8  & 0.00188 & 0.34 & 0.526 &   1.14 &  48.0 & 0.838 &    1.97 & 613 \\ 
9* & 0.00289 & 0.37 & 3.816 & 229.74 & 586.1 & 0.995 &--226.90 & 589 \\
\hline
tot& 0.00033 & 0.27 & 0.281 &   1.28 &  55.5 & 0.568 &    1.89 & 603 \\ 
$\sigma$  &  & 0.02 & 0.078 &   0.30 &  20.1 & 0.071 &    0.33 &  41 \\
\hline
\end{tabular}
\caption{Electron transfer -- shift.}
\end{table}

\begin{table}
\centering
\begin{tabular}{lcccccccccccc}
fit& RSS/TSS& $a_1$& $t_1$ & $a_2$& $t_2$ & $c_2$ & $a_4$ & $\tau$  \\
\hline
0  & 0.0164 & 1.63 & 0.122 & 3.28 &  63.9 & 0.662 &  3.16 & 784 \\ 
1  & 0.0146 & 1.37 & 0.320 & 4.13 & 104.4 & 0.779 &  2.42 & 998 \\ 
2* & 0.0126 & 1.61 & 0.125 &10.14 & 328.4 & 0.607 &--3.11 & 680 \\ 
3  & 0.0171 & 1.48 & 0.095 & 2.36 &  27.6 & 0.576 &  4.58 & 642 \\ 
4  & 0.0162 & 1.19 & 0.241 & 2.98 &  55.7 & 0.679 &  3.78 & 696 \\ 
5  & 0.0154 & 1.43 & 0.069 & 2.72 &  34.8 & 0.422 &  4.11 & 450 \\ 
6  & 0.0159 & 1.33 & 0.142 & 4.99 & 205.8 & 0.491 &  2.37 & 811 \\ 
7  & 0.0144 & 1.38 & 0.257 & 1.99 &  29.2 & 0.942 &  4.60 & 545 \\
8  & 0.0162 & 1.32 & 0.213 & 2.97 &  45.6 & 0.865 &  3.77 & 788 \\
9* & 0.0137 & 1.57 & 0.107 & 8.36 & 296.6 & 0.608 &--1.56 & 639 \\
\hline
tot& 0.0019 & 1.48 & 0.133 & 3.28 &  61.4 & 0.633 &  3.53 & 685 \\
$\sigma$ &  & 0.13 & 0.088 & 0.97 &  59.9 & 0.180 &  0.88 & 171 \\
\hline
\end{tabular}
\caption{Electron transfer -- tilt.}
\end{table}

\clearpage
\newpage

\subsection{Relaxation Plots}

\begin{figure}[hb]
\centering
\includegraphics{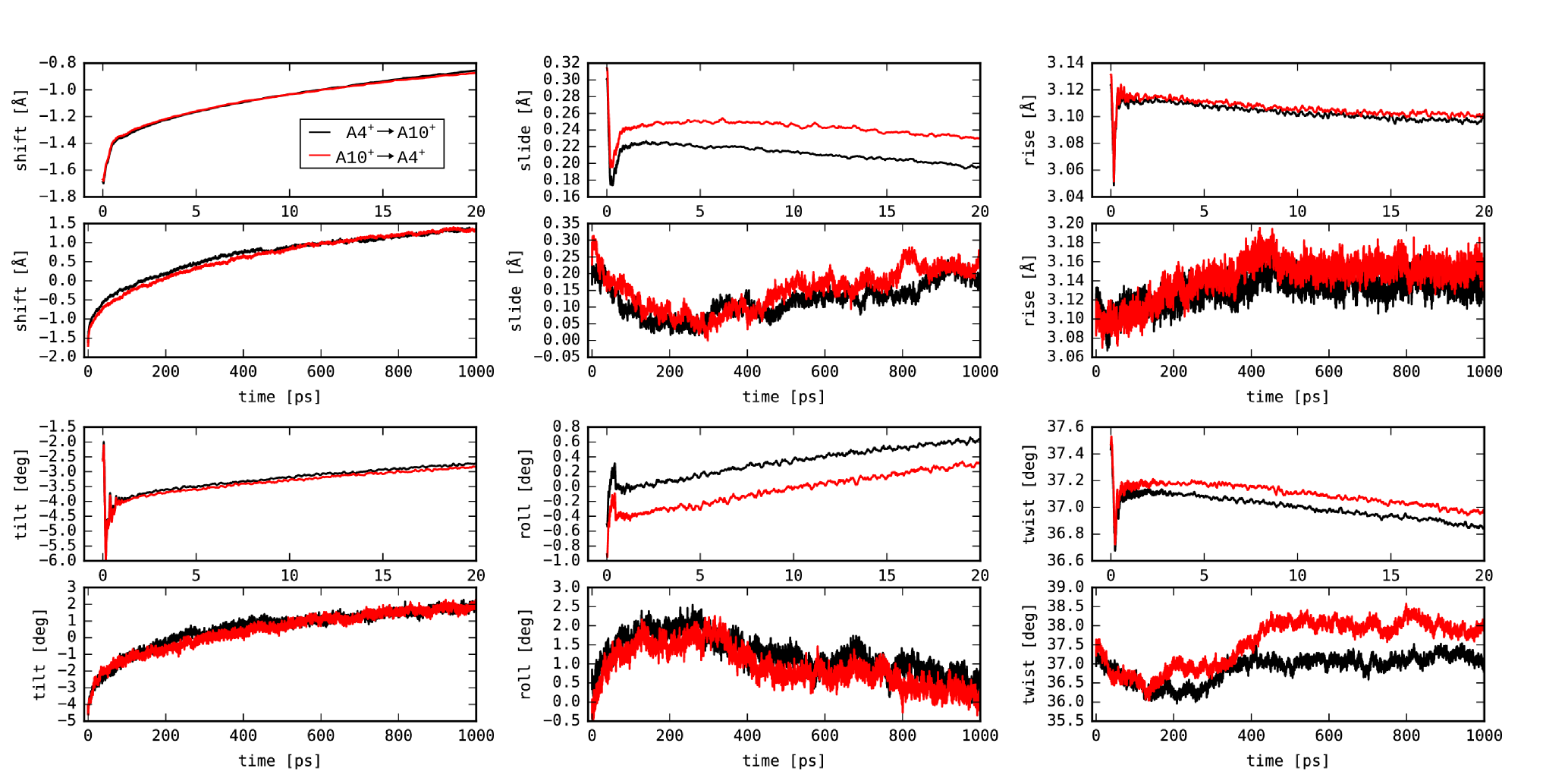}
\caption{Time evolution of the base-pair step parameters 
following the electron transfer. Note that the actual values
of shift and tilt for A$_4$- and A$_{10}$-involving reactions 
have opposite sign owing to the symmetry of the DNA sequence
and anti-symmetry of the respective parameters. For clarity,
the values of parameters involving A$_{10}$ were inverted before
plotting.}
\label{fig:relaxq}
\end{figure}

\begin{figure}
\centering
\includegraphics{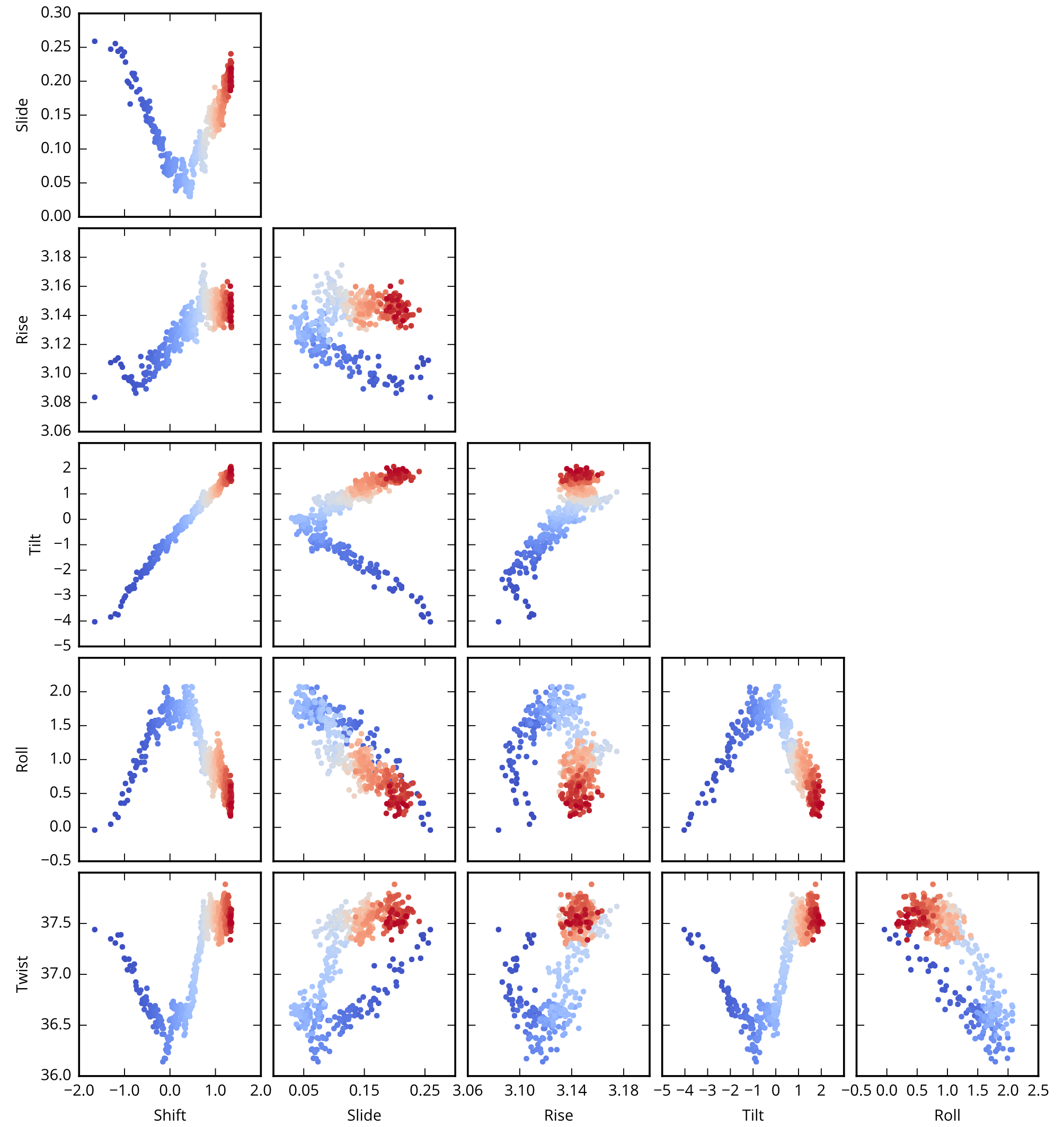}
\caption{Correlation diagrams of all possible pairs of helical
parameters of the central TA base-pair step with the adenine in
the radical cation state. The time is color
coded from blue to red.}
\end{figure}

\begin{figure}
\centering
\includegraphics{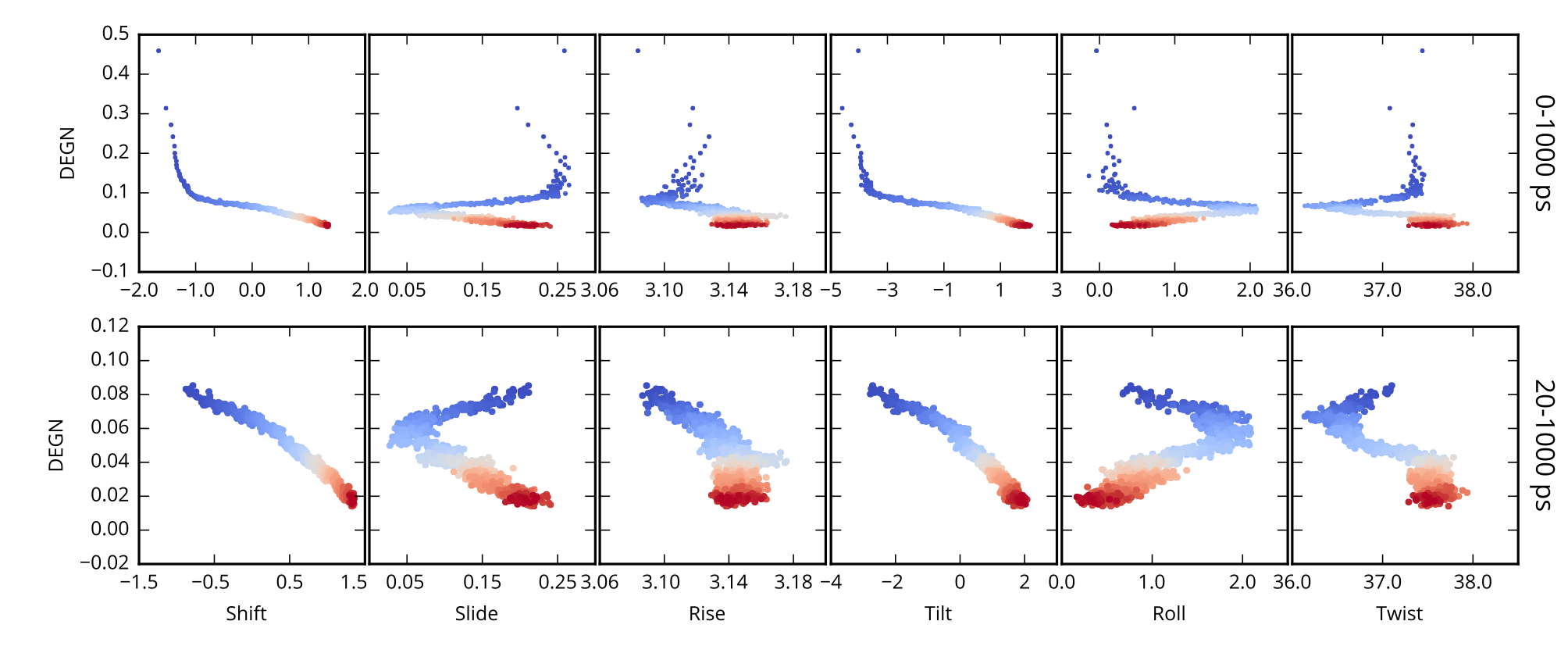}
\caption{Correlation diagrams of DEGN with all of the helical
parameters of the central TA base-pair step with the adenine in
the radical cation state. The time is color
coded from blue to red. The plots in the bottom row start at the time of 20~ps.}
\end{figure}

\clearpage
\newpage

\subsection{Relaxation simulations extended to 10 ns}

\begin{figure}[h]
\includegraphics{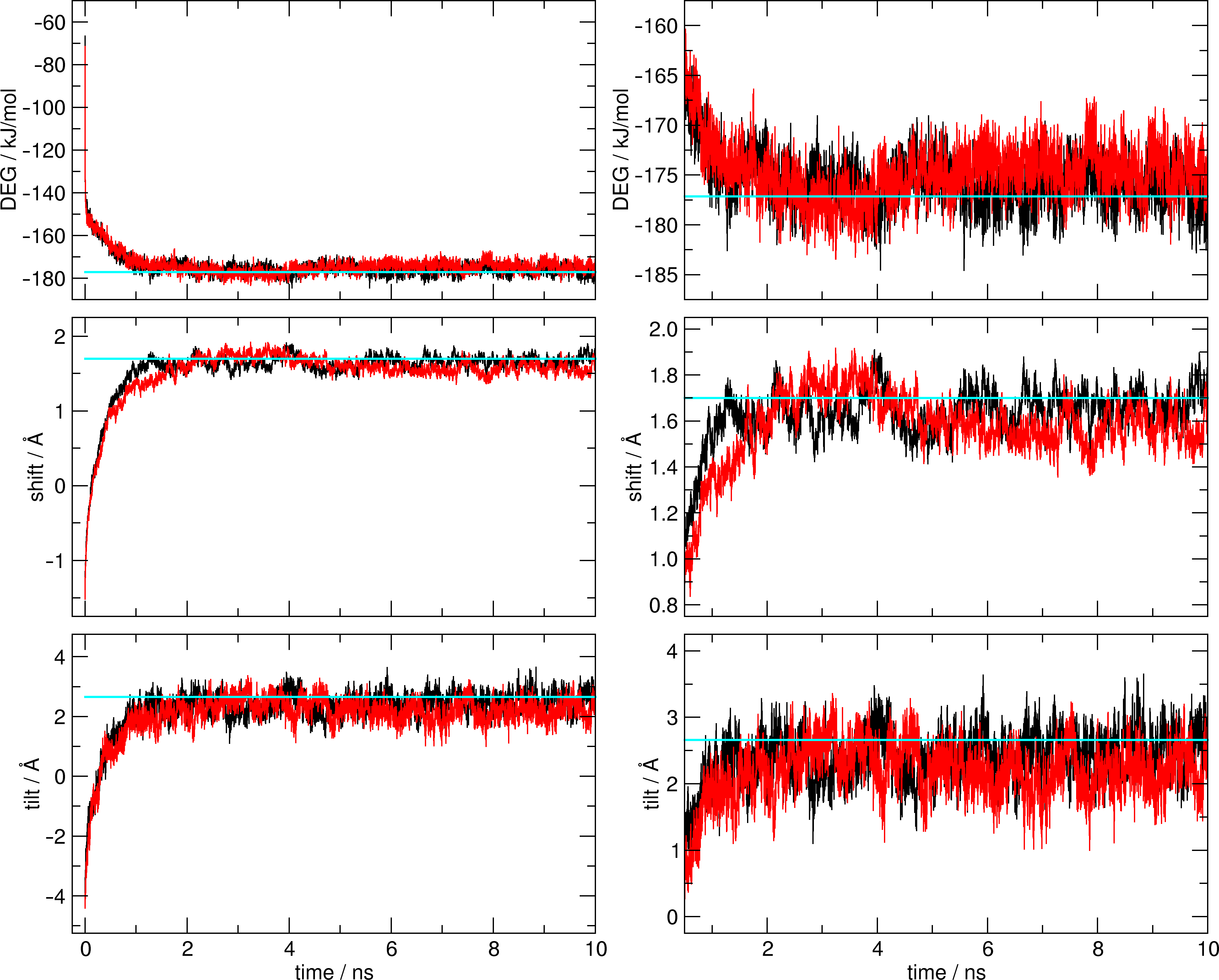}
\caption{Relaxation of DEGN (top), shift (center) and tilt (bottom)
following the electron transfer reactions
$\mathrm{A}_4^+ \to \mathrm{A}_{10}^+$ (black) and
$\mathrm{A}_{10}^+ \to \mathrm{A}_4^+$ (red;
sign was inverted before plotting for shift and tilt).
The cyan line depicts the equilibrium value of the final state.}
\end{figure}

\clearpage
\newpage

\section{Miscellaneous}

\subsection{Full author list of references truncated in the main text}
Ref.~51 -- Case \etal{} is \cite{SIAmber14};
Ref.~61 -- Olson \etal{} is \cite{SIOlson01};
Ref.~80 -- Ivani \etal{} is \cite{SIIvani16};
Ref.~82 -- Pasi \etal{} is \cite{SIPasi14}.

\end{document}